\documentclass[twocolumn,aps,prd,amsmath,amssymb,preprintnumbers]{revtex4}

\DeclareMathOperator{\sech}{sech}
\usepackage{tikz}
\usetikzlibrary{shapes.geometric}

\begin{document}

\title{Euclideanization without Complexification of the Spacetime}

\author{Nicole Drew}
\email{nicoledrew@psu.edu}
\affiliation{Institute for Gravitation and the Cosmos,
The Pennsylvania State University, 104 Davey Lab, 251 Pollock Road, University Park,
PA 16802, USA}
\affiliation{Department of Physics, The Pennsylvania State
University,
N-213 Millennium Science Complex, University Park, PA 16802, USA}

\author{Venkatraman Gopalan}
\email{vxg@psu.edu}
\affiliation{Departments of Physics \& Materials Science and Engineering, The Pennsylvania State
University,
N-212 Millennium Science Complex, University Park, PA 16802, USA}

\author{Martin Bojowald}
\email{bojowald@psu.edu}
\affiliation{Institute for Gravitation and the Cosmos,
The Pennsylvania State University, 104 Davey Lab, 251 Pollock Road, University Park,
PA 16802, USA}

\begin{abstract}
  Minkowski spacetime can be mapped by a series of projections in a
  higher-dimensional spacetime to a Euclidean space, constituting a process of
  Euclideanization shown here in detail for two dimensions. The result allows
  regularizations and computations of integrals that appear in quantum field
  theory (QFT) without performing the standard Wick rotation of time to
  imaginary values. However, there is no physical spacetime transformation
  that produces a Wick rotation. In avoiding this complexification process,
  the new Euclidenization procedure has important advantages in the
  transformations of the action principles, including fermionic fields and
  theories at constant chemical potential. In all cases, complex-valued
  amplitudes of the form $\exp(iS/\hbar)$ are mapped to real statistical
  weights $\exp(-S_{\rm E}/\hbar)$ with a Euclidean action $S_{\rm
    E}$. The procedure is also amenable to fields on curved background
  spacetimes as well as gravitational interactions.
\end{abstract}

\maketitle

\section{Introduction}

Fundamental theories of physics are formulated for fields on 4-dimensional
Minkowski spacetime (MS). The geometry of MS is closely related to properties
of hyperbolae, replacing the basic circles used in Euclidean geometry: events
at a constant spacetime distance, $\zeta$, from the origin follow surfaces of
hyperboloids, given by $r^2-t^2=\pm \zeta^2$ where $r$ is the 3-dimensional
spatial distance from the origin, $t$ is time (assuming the speed of light in
vacuum, $c=1$) and the positive (or negative) sign indicates a spacelike (or
timelike) relationship. Important restrictions on the propagation of fields
are placed by the causal structure determined by light cones in this geometry,
defined by $\zeta^2=0$. However, there are closely related technical
complications because invariant subsets in this geometry, given by
hyperboloids as analogs of Euclidean spheres, are non-compact and often lead
to divergent spacetime integrals of functions relevant in quantum field
theory. The sign changes in actions or Hamiltonians compared with Euclidean
expressions can imply additional instabilities. For these reasons, the
dynamics of quantum field theories are often evaluated not in their original
Minkowskian form, but in a Euclidean form given by the 4-dimensional
hypersphere, $r^2+t^2=\zeta^2$. The latter is achieved by a Wick rotation,
$t \rightarrow it$, that introduces imaginary time in order to replace
Minkowski spacetime with 4-dimensional Euclidean space; see for instance \cite{QFT}.

Applying a Wick rotation is an established and often
  successful procedure that in many cases can be justified independently, for
  instance in the derivation of tunneling properties via instanton solutions \cite{Instantons}.
  However, it also comes with its own disadvantages. First, it is hard to
find physical interpretations of expressions based on imaginary values that
have been assigned to observables, such as time; and space and time are
treated differently in this way. Secondly, its application in the
regularization of divergent Minkowski integrals is not always
justified by independent considerations.  Thirdly, while a Wick rotation
fulfills the technical purpose of transforming a complex amplitude to a real
statistical weight for standard action principles, such as the Klein--Gordon
field or Yang--Mills fields, it is non-unique in some cases (notably for
fermionic theories) and in others, requires further imaginary observables
(such as quantum field theories at a constant chemical potential
\cite{SignProblem,FiniteDensityQCD}). Finally, a major challenge is the
question of how to generalize the complexification procedure to quantum field
theories on non-static curved backgrounds, or to non-perturbative
gravitational interactions. In both cases, the fields in general include
metric components depending on time, which would be turned into complex
functions by a Wick rotation.

It is therefore important to study properties of Euclideanization that may be
obtained without using complex coordinates. Our work is motivated by the
transformation between Mikowskian and Euclidean geometries proposed in
\cite{RBS}. The mathematical description of this transformation is succinctly
presented in Appendix~\ref{sec:RBS}. Here we propose a new geometrical method
to visualize this transformation based on projections between the crucial
geometrical objects in Minkowskian and Euclidean geometries, given by
hyperbolae and circles, respectively.  Such projections are not unique for
non-unit curves of equal distance from the origin, but we will show that the
introduction of an auxiliary 3-dimensional Minkowski spacetime removes several
ambiguities in such a mapping. The specific projections are then designed so
as to respect the initial causal structure of Minkowski spacetime. These
constructions pinpoint the crucial place at which non-compact hyperbolae are
mapped to compact circle segments, with implications for the convergence of
spacetime integrals as they appear often in quantum field theories. We present
detailed steps of this construction in Section~\ref{sec:Projection} for
individual points in 2-dimensional Minkowski spacetime, mapped to points in
2-dimensional Euclidean space. Section~\ref{sec:Vectors} then extends this
mapping to tangent vectors and transports the metric, implementing the crucial
change of signature by an intermediate geometrical construction. The entire
construction will treat time and space on an equal footing. Performing
QFT-type integrals and their
regularization with our mapping is illustrated by an explicit example in Section~\ref{sec:Integrals}.\\
\indent The resulting mapping of metrics, summarized in
Section~\ref{sec:Mapping} and directly applied to a representative set of
field theories in Section~\ref{s:App}, is close to the results of a Wick
rotation, but it is more specific and results in an unambiguous transformation
of metric terms in a general action principle. For fermionic theories,
however, inherent ambiguities so far remain also in our formulation because a
spinor field may be transformed in different ways along with the metric
transformation.  Our method has clear advantages for field theories at
constant chemical potential, where the traditional Wick rotation results in a
complex action when applied for instance to a scalar field whose Klein--Gordon
current implies a density linear in time derivatives. In contrast, our
transformation keeps the density real and only multiplies the volume element
by the single factor of $i$ that is needed to turn the amplitude
$\exp(iS/\hbar)$ into a statistical weight, $\exp(-S_{\rm E}/\hbar)$ with a
Euclidean action $S_{\rm E}$ replacing the Minkowskian $S$. In combination
with an older proposal for a specific spinor transformation
\cite{FermionWick}, both bosonic and fermionic theories at constant chemical
potential are mapped to real Euclidean actions, suggesting a solution to the
sign problem of importance especially in lattice QCD
\cite{SignProblem,FiniteDensityQCD}.  Finally, we will demonstrate that all
ingredients of our geometrical construction can, at least in principle, be
generalized to field theories on non-static curved backgrounds and
non-perturbative gravity.

\section{Conic projection}
\label{sec:Projection}

In this main section, we construct a mapping from the 2-dimensional
Minkowski spacetime to a Euclidean plane, respecting important geometrical
relationships. Basic objects in these two geometries are curves of unit
distance from the origin, given by unit hyperbolae (both spacelike and
timelike) and the unit circle, respectively. Any geometrical mapping should
therefore map unit hyperbolae to the unit circle. However, this basic
condition does not provide a unique answer as to how the mapping should be
defined for points not on one of the unit hyperbolae, or how a non-unit
hyperbola should be mapped to a non-unit circle. We will show that this
problem can be solved by using an intermediate extension of 2-dimensional
Minkowski spacetime to a specific 3-dimensional Minkowski spacetime, together
with a mechanism to lift points and directions into the 3-dimensional
extension. The same construction will then be seen to solve also the main
question of how to change the signature of the 2-dimensional geometry from
Minkowskian to Euclidean without introducing complex coordinates or metric
components. Our main conditions for the construction of such a
  procedure are (i) that it reproduce results compatible with a Wick rotation
  in well-understood cases of field theories on Minkowski spacetime, and (ii)
  that it be generalizable to situations in which a Wick rotation or
  alternatives remain underexplored, such as curved spacetime or dynamical
  gravity.

Anticipating the result of our construction for 2-dimensional
  Minkowski spacetime, the mapping is easy to state and takes the form
  \[
    (x,t)\mapsto \sqrt{\frac{|x^2-t^2|}{x^2+t^2}}(x,t)=(\bar{x},\bar{t})
  \]
  as summarized in Eq.~(\ref{eqn:Complete-Projection-p0-to-pBar}) below.
  In regions that are timelike or spacelike related to the origin, the mapping
  confirms the expectation that hyperbolae are mapped to circles.
  The mapping on lightlike related regions in the Mikowski space is obtained as a limiting case from
  our construction and sends all these events to the Euclidean
  origin. (Depending on the pole structure on the light cone for a given
  theory, the mapping may have to be modified by removing continuity toward
  the light cone. We leave this question open for further studies.) Conversely, the inverse of the above transformation maps all the finite points on the diagonals in the Euclidean construction to the four infinity poles of the Minkowski lightlines.

  The line elements in the mapped coordinates are given by
  ${\rm d}x^2-{\rm d}t^2$ and ${\rm d}\bar{x}^2+{\rm d}\bar{t}^2$,
  respectively. In applications to quantum field theory, the point mapping is
  directly applied to positions at which the fields are evaluated, and the
  changed line element implies specific sign factors wherever the metric is
  used in the integration measure or to raise and lower indices. In general, a
  mapping of this form is far from being unique because it cannot be an
  isometry or subject to other restrictive mathematical conditions. For this
  reason, we require that the mapping be derived from a more detailed
  geometrical construction that clarifies the change of signature and fulfills
  our two physical conditions stated above, namely, reproducing established
  results in known cases and being generalizable to theories in which
  spacetime curvature is essential. We also show in Appendix~\ref{sec:RBS}
  that the above transformation is but one special case of a more general
  transformation discussed in ~\cite{RBS}; thus many more mappings are
  possible within the geometry to be discussed next.  We now embark on our
  discussion of the geometrical construction and return the the physical
  conditions in Sections~\ref{sec:Integrals} and \ref{s:App}.

\subsection{Basic setup}

\begin{figure*}[htbp]
    \centering
    \includegraphics[width=16cm]{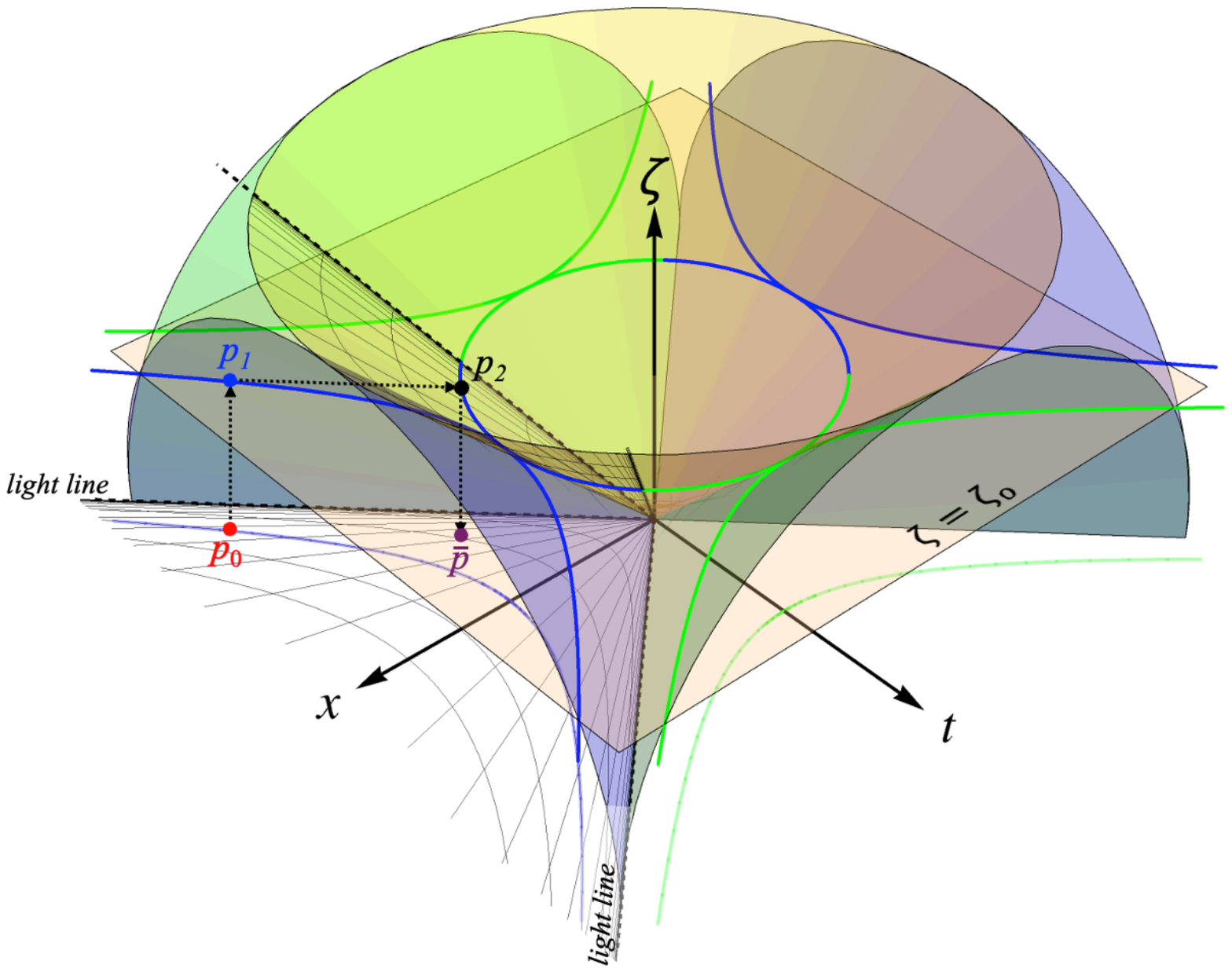}
    \caption{A 3D Minkowski spacetime, $(x,t,\zeta)$, where $\zeta$ is the
      spacetime length defined by $x^2-t^2=\mp \zeta^2$; the negative
      (positive) value of $\zeta^2$ corresponds to timelike (spacelike)
      events. The $xt$-plane at $\zeta=0$ is the original 2D Minkowski
      plane. A central cone (in yellow) defined by
      Eq.~(\ref{eqn:Central-Cone}) and four half-cones defined by
      Eq.~(\ref{eqn:Half-Cones}) are shown. The blue (green) half-cones are
      over the spacelike (timelike) events of the Minkowski plane. A beige
      colored plane at $\zeta = \zeta_o > 0$ is shown intersecting the
      cones. The intersection of the plane, $\zeta = \zeta_o$, with the
      central yellow cone forms arcs of a circle defined by
      $x^2 -\zeta_o^2+t^2=0$. The intersection of the plane,
      $\zeta = \zeta_o$, with the four half-cones forms the four branches of
      the hyperbolae defined by $x^2 \pm \zeta_o^2-t^2=0$. Iso-lines of
      constant spacetime length, $\zeta$ and of constant hyperbolic angle,
      $\sigma$ are shown in dull grey for one of the four quadrants on the
      Minkowski plane. The dashed black lines in the $xt$-plane are the light
      lines as indicated. The point transformations
      $p_0 \rightarrow p_1 \rightarrow p_2 \rightarrow \Bar{p}$ are depicted,
      where $p_0$ and $\Bar{p}$ lie on the $\zeta=0$ plane and the points
      $p_1$ and $p_2$ lie on the $\zeta=\zeta_o >0$ plane. The speed $c=1$ is
      assumed.}
    \label{fig:conicproj}
\end{figure*}

Consider the 3-dimensional $(x,t,\zeta)$ space depicted in
Fig.~\ref{fig:conicproj}, where $x$ is a space coordinate along which relative
motion between two inertial frames occurs. Further, $t$, is the time
coordinate which as usual, can be read as a distance, $ct$, assuming the speed
of light in vacuum, $c=1$. The $xt$-plane at $\zeta=0$ is the original
Minkowski spacetime. The extension to an auxiliary third dimension allows us
to introduce additional structures that help us to define different components
of the mapping to the Euclidean plane.

Our conical construction comprises of a
central cone with its conical axis parallel to the $\zeta$ axis, given by
\begin{equation} \label{eqn:Central-Cone}
    x^2 - \zeta^2 + t^2 = 0, \quad  \zeta>0    \,.
\end{equation}
Abutting this central cone are four half-cones, each lying over one full
spacelike or timelike quadrant of the Minkowski space. Their surfaces are defined by
\begin{equation} \label{eqn:Half-Cones}
    x^2 \pm \zeta^2 - t^2 = 0, \quad \zeta>0    
\end{equation}
where the plus sign for the $\zeta^2$ term is for timelike half cones, and
the minus sign is for the spacelike half cones. The edges of the half cones
in Fig.~\ref{fig:conicproj} come into contact with each other and the
Minkowski plane precisely on the light lines in the $xt$-plane defined by
$x=\pm t$. The central cone touches the spacelike cones along the lines
defined by $\zeta=\pm x$, and the timelike cones along the lines
$\zeta = \pm t$.

\begin{figure}[!h]
    \centering
    \includegraphics[width=10cm]{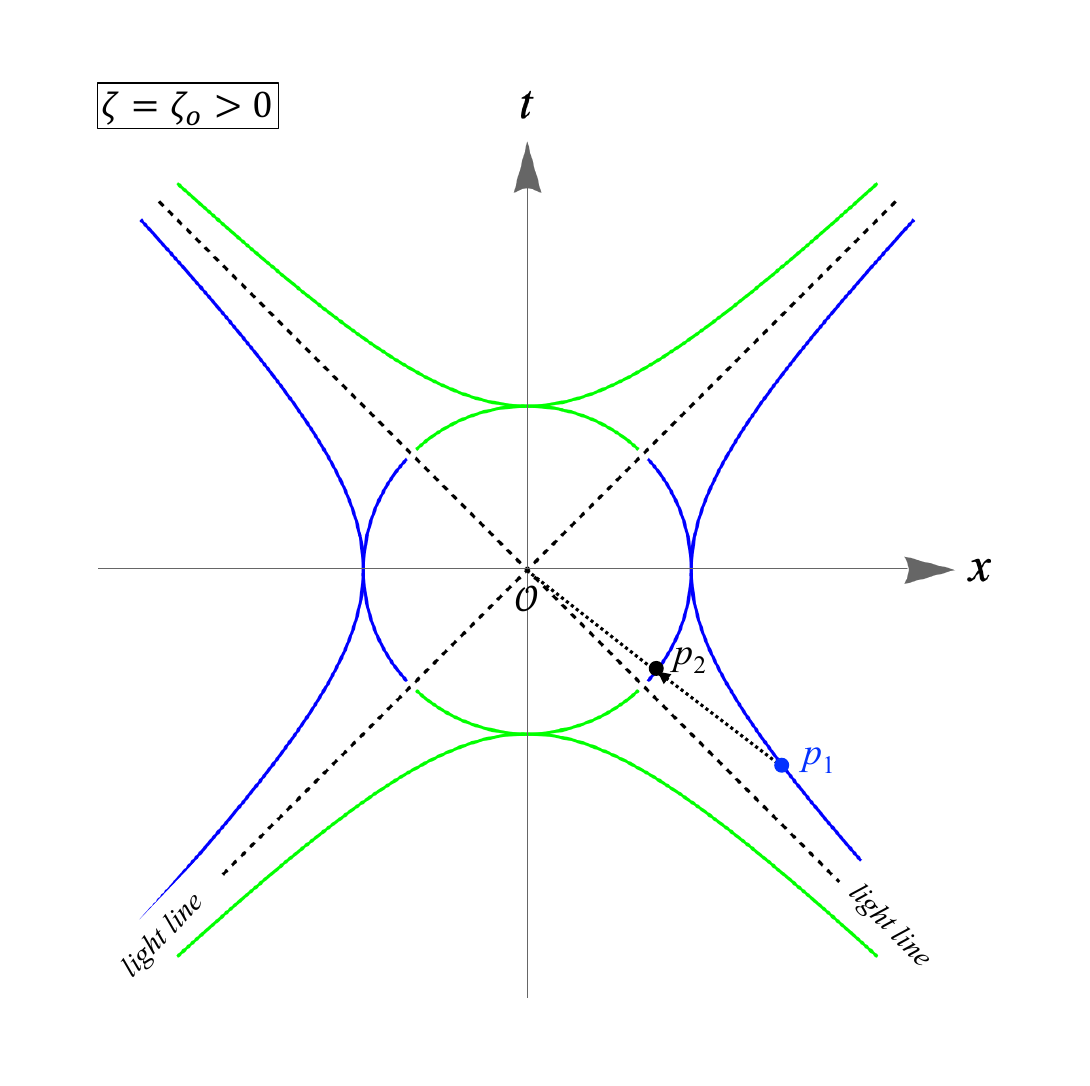}
    \caption{A $\zeta = \zeta_o > 0$ plane that cuts the 3-D Minkowski spacetime depicted in Fig.~\ref{fig:conicproj} reveals the circular arc segments of Euclidean radius $\zeta_o$ of the conical section of the central cone, and the hyperbolic branches of hyperbolic radius $\zeta_o$ formed by the conical sectioning of the adjacent half cones. The blue (green) conical sections represent spacelike (timelike) events. A mapping, $p_1 \rightarrow p_2$, of a general point $p_1$ on a hyperbolic branch to a point $p_2$ on the circle is made by the parallel projection along the sectioning plane in the radial direction marked by the dotted black line with an arrow. Light lines are depicted as dashed black lines.}
    \label{fig:Circ&Hyperbolae}
\end{figure}

If we section these cones with a plane parallel to the $xt$-plane at
$\zeta = \zeta_o$, the cross section of this plane is depicted in
Fig.~\ref{fig:Circ&Hyperbolae}. The four branches of a hyperbola at a
hyperbolic radius of $\zeta_o$ from the origin of the plane, and a circle of
Euclidean radius $\zeta_o$ from the origin of the plane can be seen. This
construction allows us to map a Minkowski point anywhere, not just on a unit
hyperbola, to a point on a non-unit circle by first lifting it to the relevant
half-cone and mapping it from there to the central cone, where its image will
lie on a circle around the $\zeta$-axis. We will show in detail how such a
mapping can be constructed by parallel projection in the first step, and
central projection along radial lines in the second step. Another parallel
projection can then be used to map the circles back to the $xt$-plane, now
equipped with Euclidean geometry. Because these steps do not use specific
properties such as right angles in the 3-dimensional geometry, they are able
to bridge between different geometries.

A further observation shows where the transition from Minkowski to Euclidean
signature is performed using the conical construction. On each point
$(x,t,\zeta)$ on a half cone, $\zeta^2$ equals the spacetime interval from the
origin to $(x,t)$, and $\zeta$ is the spacetime length. On the central cone,
$\zeta$ equals the Euclidean distance from the origin to $(x,t)$. The crucial
step is therefore the transition between the cones. Our 3-dimensional
extension allows us to define this transition by flipping metric components,
$g_{tt}$ and $g_{\zeta\zeta}$ on timelike half-cones, and $g_{xx}$ and
$g_{\zeta\zeta}$ on spacelike half-cones. In the latter case, we initially
obtain a negative Euclidean line element, $-{\rm d}x^2-{\rm d}t^2$, which we
may easily map to the usual positive version by a conformal transformation,
multiplying the whole line element by negative one. (We note that flipping the
metric components can be avoided if we interpret the $x\zeta$-plane in
timelike quadrants (at a fixed $t=0$) and the $t\zeta$-plane in spacelike
quadrants (at a fixed $x=0$) as the corresponding Euclidean planes within a
fixed auxiliary manifold equipped with two different line elements,
${\rm d} x^2-{\rm d}t^2\pm{\rm d}\zeta^2$, where the plus(minus) signs
correspond to $x\zeta$ ($t\zeta$) planes. This sign choice implies that both
planes, given by $t=0$ in the geometry ${\rm d} x^2-{\rm d}t^2+{\rm d}\zeta^2$
and by $x=0$ in the geometry ${\rm d} x^2-{\rm d}t^2-{\rm d}\zeta^2$,
respectively, are spacelike, although with the negative-Euclidean sense for
the latter case.  Using these different planes for the two cases of quadrants
then replaces the step of flipping metric components. However, we prefer to
work with a single Euclidean plane for all quadrants, and therefore flip
metric components as described.)

All sectors of Minkowski spacetime will thereby be mapped to corresponding
regions of the Euclidean plane, treating space and time on an equal
footing. The final result of this mapping resembles what can be accomplished
by a Wick rotation, but it is obtained here without using complex numbers for
coordinates or metric components. This property will be beneficial in
applications to quantum field theory as shown in Sec.~\ref{s:App}. As another
key advantage, we will see that the ingredients of our new procedure can be
generalized to curved spacetimes.

\subsection{Point projections in Cartesian coordinates}

\begin{figure*}[htbp]
    \centering
    \includegraphics[width=16cm]{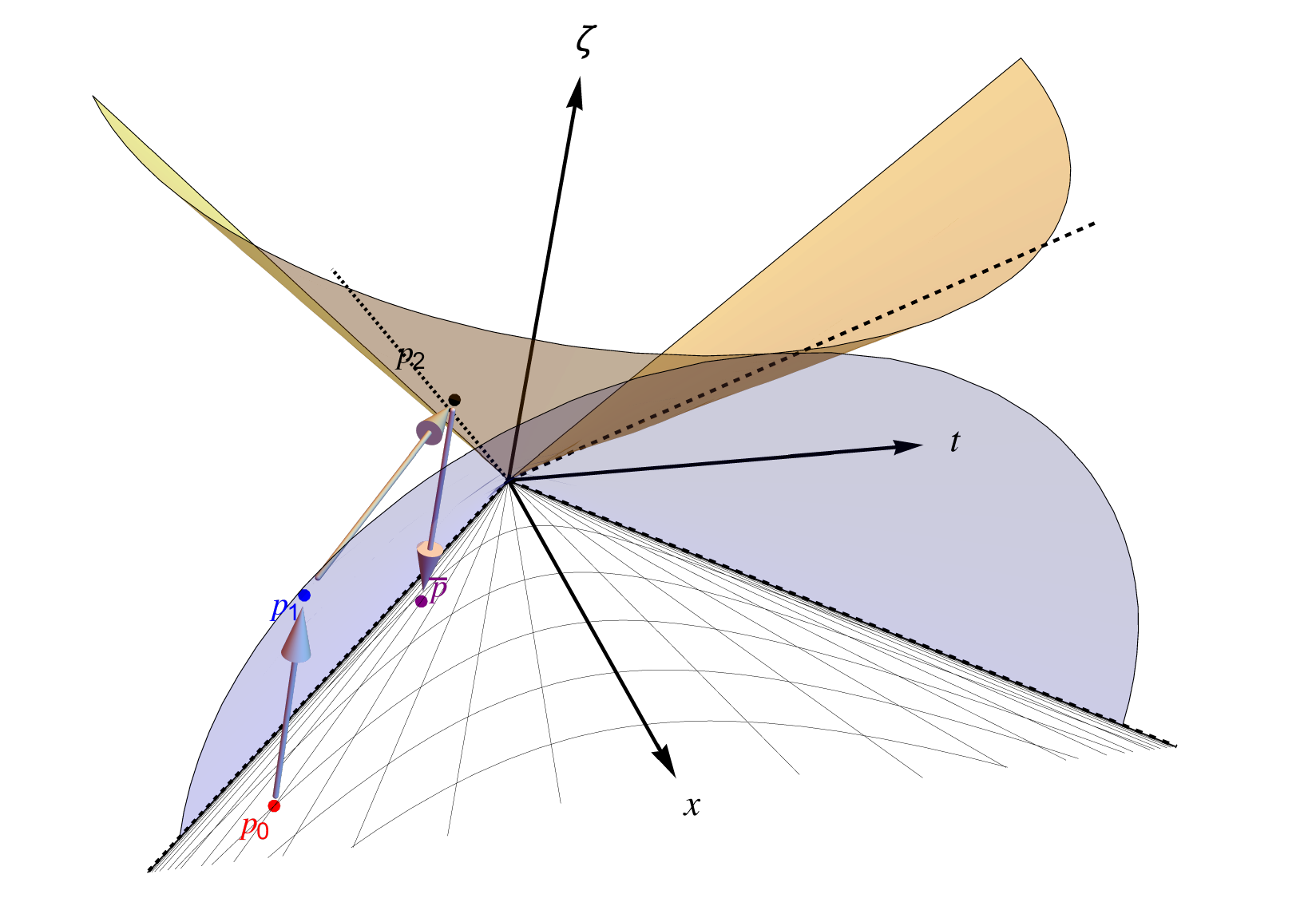}
    \caption{An overview of the three projections, $p_0 \rightarrow p_1
      \rightarrow p_2 \rightarrow \Bar{p}$. A general point, $p_0$, is
      projected from the hyperbolic Minkowski plane at $\zeta=0$ up to a
      unique point, $p_1$, on the surface of the corresponding half-cone
      (\textit{first projection}), then radially inwards parallel to the
      Minkowski plane to a unique point $p_2$ on the surface of the central
      cone (\textit{second projection}, see Fig.~\ref{fig:Circ&Hyperbolae})
      followed by a final projection back down to a unique point, $\Bar{p}$,
      on the 2D Minkowski plane at $\zeta=0$ (\textit{third projection}), with
      the resultant geometry becoming Euclidean by a flip of metric components
      without the need for invoking Wick's rotation of the time
      coordinate. These projections are depicted in detail in the figures that
      follow. The net transformation in the Minkowski $xt$-plane, after
      suppressing the third coordinate, is given by $p_0\colon (x_0,t_0)
      \rightarrow  \bar{p}\colon \left(\Omega x_0 ,\Omega t_0 \right)$. The factor $\Omega$ is given by Eq.~(\ref{eqn:Scaling_Factor_Omega_as_a_f(xo, to)}).}
    \label{fig:Allprojs}
\end{figure*}

We now perform an explicit series of point projections following our general
procedure. The general idea is captured by Figures~\ref{fig:conicproj} and
\ref{fig:Allprojs}. Each of these projection steps is described in detail
next. Starting with a point $p_0=(x_0,t_0)$ on the Minkowski plane, we embed
it in our 3-dimensional auxiliary space as $(x_0, t_0, 0)$, project it on the
light cone of its Minkowski sector and from there to the central cone and back
to the hypersurface of $\zeta=0$ with image $(\Bar{x},
\Bar{t},0)$. Interpreting this hypersurface now as a Euclidean plane, the
final Euclidean point is $\Bar{p}=(\Bar{x}, \Bar{t})$.

This procedure will be performed in three projection
steps: \textit{First projection}, from the original quadrant onto the surface
of its corresponding half-cone; \textit{Second projection}, from this
half-cone onto the surface of the central cone; and \textit{Third projection}, projecting this point back on to a Euclidean $\Bar{x}\Bar{t}$-plane parallel to the original Minkowski $xt$-plane. 

\subsubsection{First projection: Minkowski plane to light cone} \label{sec:First_Projection}

\begin{figure*}[htbp]
    \centering
    \includegraphics[width=16cm]{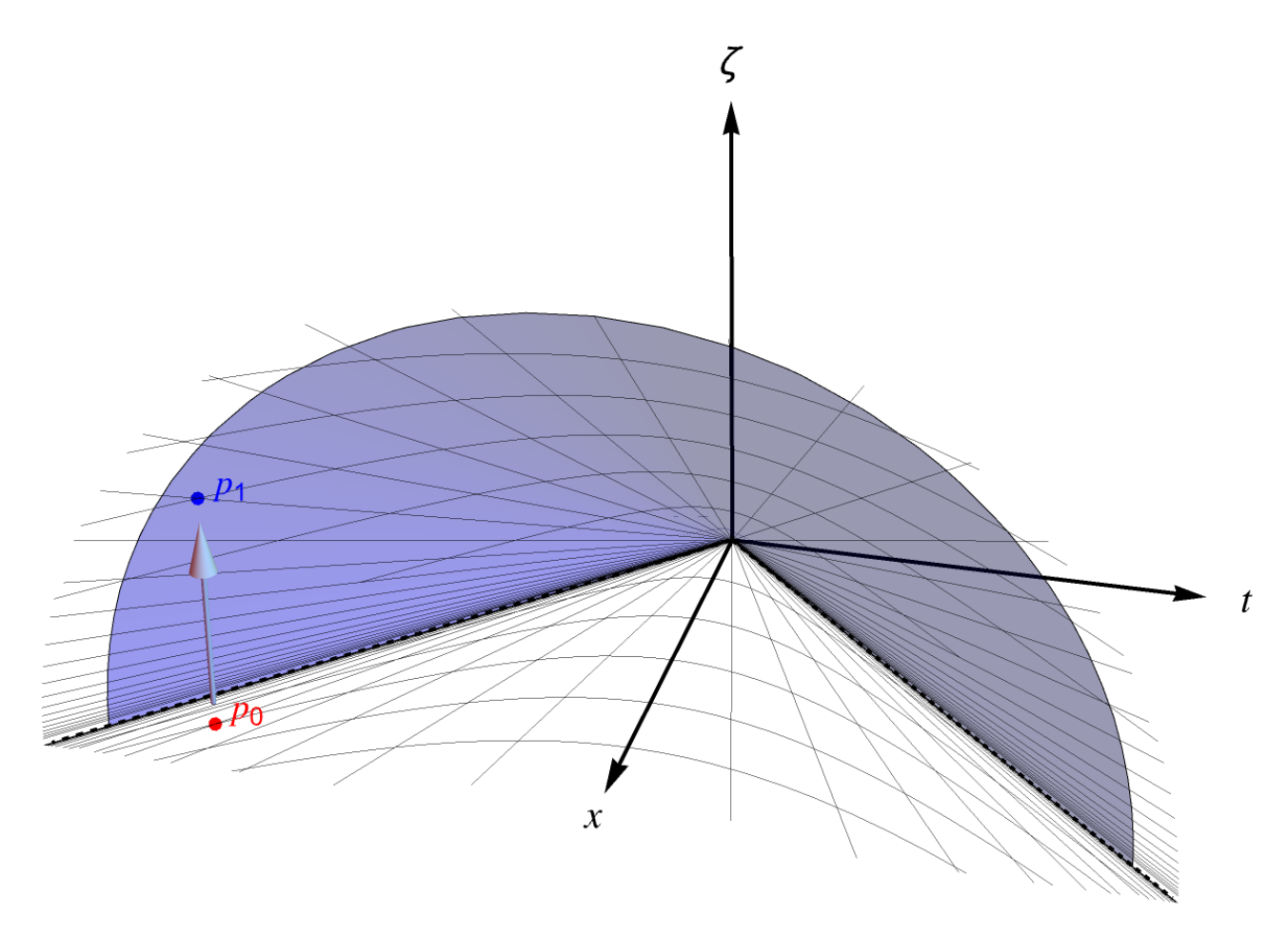}
    \caption{Illustration of the \textit{first projection} of a point
      $p_0\colon(x_0, t_0, 0)$ from the Minkowski plane at $\zeta=0$ to a
      point $p_1\colon (x_1, t_1, \zeta_1)=(x_0, t_0, \zeta_0=+\sqrt{|x_0^2-t_0^2|})$ on the surface of the half-cone over one of the spacelike Minkowski quadrants as shown. The point transformation is achieved by a parallel projection along the $\zeta$-axis. A similar mapping occurs for events in the other three quadrants.}
    \label{fig:1stproj}
\end{figure*}

A point $p_0=(x_0,t_0)$ in a quadrant of the original Minkowski space is first mapped to the
point, $p_1$, on the corresponding half-cone with the same $(x,t)$-coordinates, but
with a new $\zeta$-coordinate, $\zeta_0=+ \sqrt{|x_0^2-t_0^2|}$. The point
$p_0$, embedded in the auxiliary spacetime as $(x_0,t_0,0)$, is thus translated by the vector
$\textbf{d}_0=\left( 0,0,+\zeta_0 \right)$ in order to obtain the point
$p_1$. The first projection can be summarized as
\begin{equation} \label{eqn:Mapping-points-p_0-to-p_1}
    (x_0,t_0,0) \rightarrow p_1 = (x_1,t_1,\zeta_1)=\left ( x_0, t_0, \zeta_0 \right )
\end{equation}
and is visualized in Fig.~\ref{fig:1stproj}. 

\subsubsection{Second projection: Light cone to central cone} \label{sec:Second_Projection}

Consider the plane $\zeta=\zeta_0$
depicted in Fig.~\ref{fig:conicproj} with beige color. Now using central projection in order to preserve the timelike and
spacelike sectors of the Minkowski plane or its shift to $\zeta_0$, the point
$p_1$ is translated along the direction of a radial vector $\hat{\textbf{d}}_1$,
given by
\begin{equation} \label{eqn:d1-unit-vector}
    \hat{\textbf{d}}_1=\left(-x_0,-t_0,0\right)\,.
\end{equation}
For our purposes, it is sufficient to use only the direction of this vector
and ignore its length, which would depend on which geometry we assign to the $\zeta=\zeta_0$-plane. 

\begin{figure*}[htbp]
    \centering8
    \includegraphics[width=16cm]{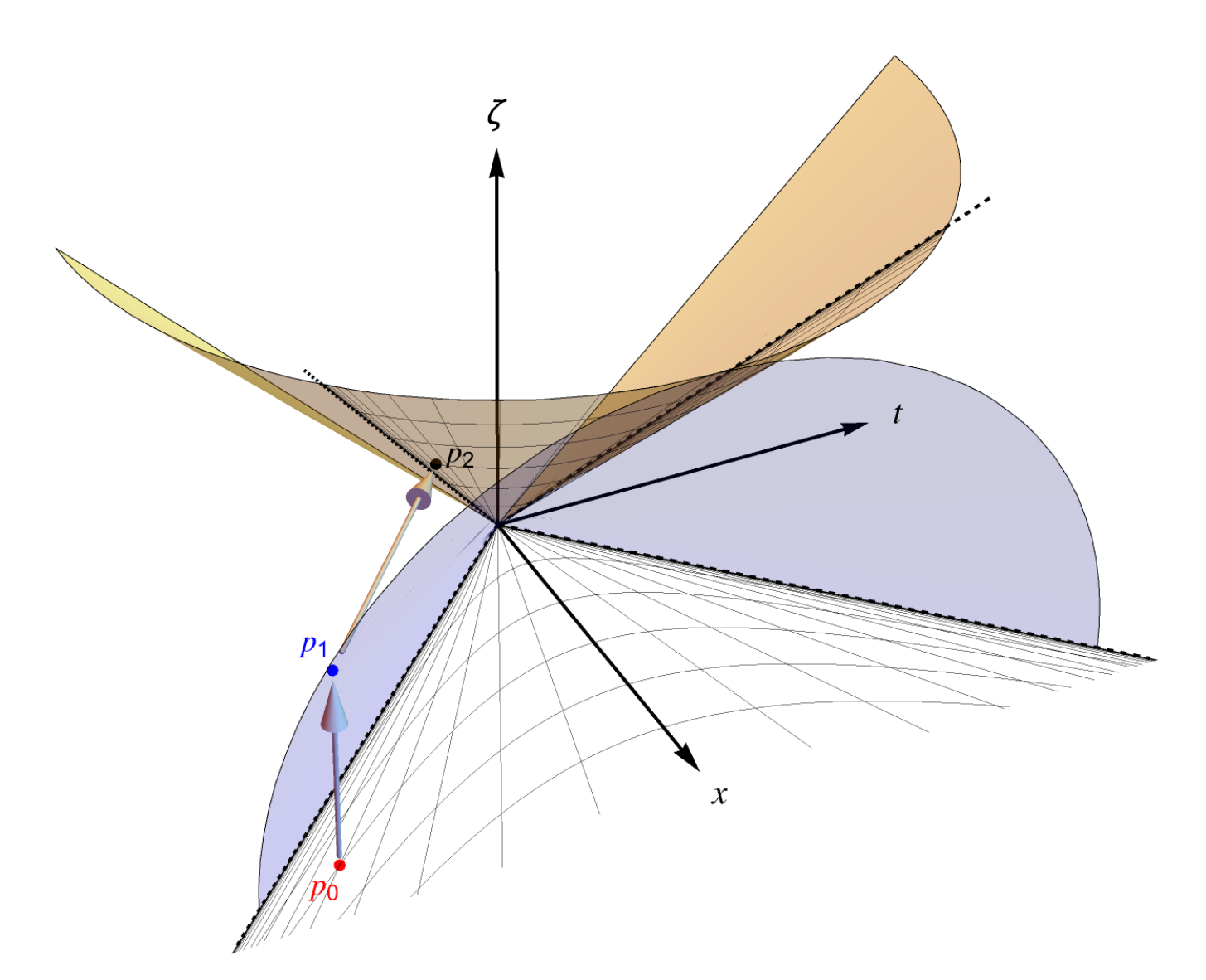}
    \caption{Illustration of the \textit{second projection} of a point $p_1:(x_1, t_1, \zeta_1)=(x_0, t_0, \zeta_0=+\sqrt{|x_0^2-t_0^2|})$ on the surface of the half-cone over one of the spacelike Minkowski quadrants to a point $p_2: (x_2,t_2,\zeta_2)=\left (\Omega x_0, \Omega t_0, \zeta_0 \right )$ on the surface of the central cone defined by Eq.~(\ref{eqn:Central-Cone}). The factor $\Omega$ is given by Eq.~(\ref{eqn:Scaling_Factor_Omega_as_a_f(xo, to)}). The point transformation is achieved by a projection parallel to the Minkowski $xt$-plane, and radially inwards towards the $\zeta$-axis. A similar mapping occurs for events in the other three quadrants that were mapped in the first projection to their corresponding half-cones.}
    \label{fig:2ndproj}
\end{figure*}

The vector $\hat{\textbf{d}}_1$ points radially inward, toward the
$\zeta$-axis, such that $p_1$ and $p_2$ have the same $\zeta$-component equal
to $\zeta_0$. The coordinates of $p_2$ are obtained by requiring it to lie
on the intersection of this plane with the central cone, given by 
the circle defined by $x_2^2+t_2^2=+\zeta_0^2$. Since $p_1$ and $p_2$ then
lie on the radial line (see Fig.~\ref{fig:Circ&Hyperbolae}) whose direction is given by
Eq.~(\ref{eqn:d1-unit-vector}), we obtain the constraint
$t_1/x_1=t_2/x_2$. Noting from Eq.~(\ref{eqn:Mapping-points-p_0-to-p_1}) that
$t_1=t_0$ and $x_1=x_0$, we can solve for the coordinates of the point $p_2$ as,
\begin{equation}
  p_2 = \left(\Omega x_0 ,\Omega t_0, \zeta_0  \right)\,,
\end{equation}
where $\Omega$ is given by
\begin{equation} \label{eqn:Scaling_Factor_Omega_as_a_f(xo, to)}
    \Omega (x_0,t_0)  = \sqrt{\frac{|x_0^2-t_0^2|}{x_0^2+t_0^2}}= \frac{\zeta_0}{r_0}
  \end{equation}
with $r_0=+\sqrt{x_0^2+t_0^2}$. The second projection can thus be summarized as
\begin{equation} \label{eqn:Mapping-points-p_1-to-p_2}
    p_1 = (x_0,t_0, \zeta_0) \rightarrow p_2 = (x_2,t_2,\zeta_2)=\left (\Omega x_0, \Omega t_0, \zeta_0 \right )
\end{equation}
and is visualized in Fig.~\ref{fig:2ndproj}.

A given half-cone and the central cone are both defined in a 3-dimensional
extension of a 2-dimensional plane, but with different sign choices in order
to obtain the desired orientation of the cones' axes: A
half-cone is obtained if the third dimension, $\zeta$, extends the 2-dimensional
Minkowski spacetime such that this coordinate appears in the metric with the
same sign as $x$ for a timelike quadrant, and with the same sign as $t$ for a
spacelike quadrant. The central cone is obtained in a 3-dimensional extension
of the Euclidean plane such that $\zeta$ appears in the metric with a sign
opposite to the two Euclidean coordinates. Geometrically, the transition
between the cones is therefore achieved by flipping the $\zeta$ and $t$ components
(timelike quadrants) and the $\zeta$ and $x$ components (spacelike quadrants),
respectively. In the final projection, removing the $\zeta$-direction, only
metric components with equal signs will remain, establishing the transition to a
Euclidean signature.

\subsubsection{Third projection: Central cone to Euclidean plane} \label{sec:Third_Projection}

\begin{figure*}[htbp]
    \centering
    \includegraphics[width=16cm]{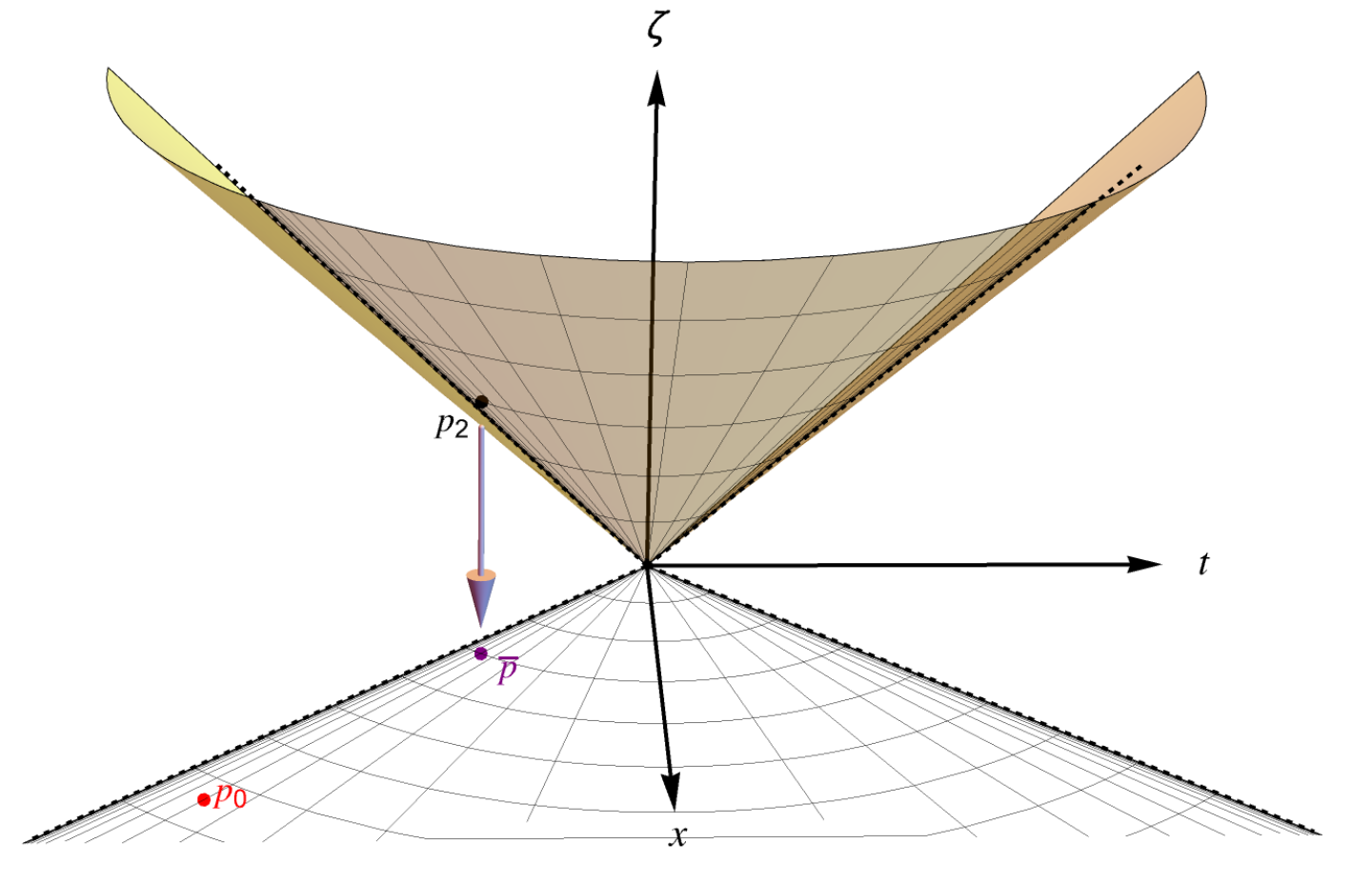}
    \caption{Illustration of the \textit{third projection} of a point $p_2: (x_2,t_2,\zeta_2)=\left (\Omega x_0, \Omega t_0, \zeta_0 \right )$ on the surface of the central cone defined by Eq.~(\ref{eqn:Central-Cone}) to the point $\bar{p} = \left(\Omega x_0 ,\Omega t_0, 0 \right)$ back on the Minkowski plane defined by $\zeta=0$. The factor $\Omega$ is given by Eq.~(\ref{eqn:Scaling_Factor_Omega_as_a_f(xo, to)}). The point transformation is achieved by a parallel projection parallel to the $-\zeta$-axis. A similar mapping occurs for events in the other three quadrants that were mapped in the second projection to the central cone.}
    \label{fig:3rdproj}
\end{figure*}

The final projection is a point transformation that maps the point $p_2$ on the central cone to the Euclidean plane with coordinates $(\Bar{x},\Bar{t})$, coincident with the original Minkowski plane with coordinates $(x,t)$, but equipped with a Euclidean metric. The point maintains its previous position with respect to the $xt$-plane, but loses its height along the $\zeta$-axis. A summary of this projection is given as
\begin{eqnarray} \label{eqn:Mapping-points-p_2-to-p_3}
    p_2 &=& (x_2,t_2,\zeta_2)=\left (\Omega x_0, \Omega t_0, \zeta_0 \right )
            \rightarrow p_3 = (x_3,t_3,\zeta_3)\nonumber\\
  &=& \left (\Omega x_0, \Omega t_0, 0 \right )
\end{eqnarray}
and is visualized in Fig.~\ref{fig:3rdproj}.

\subsubsection{Combined point projection within a plane} \label{sec:Combined_Projection}

All three projections, combined, are shown in Fig.~\ref{fig:Allprojs}. Having completed the three point projections described above, we now choose to
suppress the third coordinate corresponding to the $\zeta$-axis, and write the
transformation from points $p_0$ to $\bar{p}$ as
\begin{equation} \label{eqn:Complete-Projection-p0-to-pBar}
   p_0 = (x_0,t_0) \rightarrow  \bar{p} = \left(\Omega x_0 ,\Omega t_0 \right)
 \end{equation}
 where
   \[
     \Omega (x_0,t_0)  = \sqrt{\frac{|x_0^2-t_0^2|}{x_0^2+t_0^2}}\,.
   \]
    Note that the point $\bar{p}$ is simply the final point $p_3$ in
Eq.~(\ref{eqn:Mapping-points-p_2-to-p_3}) with its $\zeta$-component
suppressed. The point $p_0$ is the same as the original point in
Eq.~(\ref{eqn:Mapping-points-p_0-to-p_1}), but now with its $\zeta$-component
suppressed. Two-coordinates notations, $p_0$ and $\Bar{p}$, will refer henceforth to the 2-dimensional Minkowski and Euclidean planes, respectively.

\subsection{Point projections in hyperbolic coordinates} \label{sec:Point-Projections-in-Hyperbolic-Coordinates}

The radial projection in the second step, defined such that each spacelike and
timelike quadrant is preserved, makes it convenient to use a hyperbolic
version of polar coordinates.  To this end, consider a point in a spacelike
quadrant of the Minkowski $xt$-plane, whose hyperbolic coordinates are
$(\zeta_0,\sigma_0)$, such that
$p_0=(\zeta_0 \cosh{\sigma}_0,\zeta_0 \sinh{\sigma}_0)$, corresponding to the
point
\begin{equation} \label{eqn:p=xi(cosh_sigma, sinh_sigma)}
    (\zeta_0 \cosh{\sigma}_0,\zeta_0 \sinh{\sigma}_0, 0)
  \end{equation}
in our auxiliary spacetime.
The first projection corresponding to Eq.~(\ref{eqn:Mapping-points-p_0-to-p_1}) now reads
\begin{equation} \label{eqn:Polar-Mapping-points-p_0-to-p_1}
    (\zeta_0 \cosh{\sigma}_0,\zeta_0 \sinh{\sigma}_0, 0) \rightarrow p_1=(\zeta_0 \cosh{\sigma}_0,\zeta_0 \sinh{\sigma}_0, \zeta_0)\,.
\end{equation}
The second projection corresponding to Eq.~(\ref{eqn:Mapping-points-p_1-to-p_2}) now reads
\begin{eqnarray} \label{eqn:Polar-Mapping-points-p_1-to-p_2}
     &&p_1=(\zeta_0 \cosh{\sigma}_0,\zeta_0 \sinh{\sigma}_0, \zeta_0)
           \nonumber\\
 &\rightarrow & p_2  = (\Omega \zeta_0 \cosh{\sigma}_0,\Omega \zeta_0 \sinh{\sigma}_0, \zeta_0) 
\end{eqnarray}
where $\Omega $ in the hyperbolic coordinates can be derived by substituting Eq.~(\ref{eqn:p=xi(cosh_sigma, sinh_sigma)}) into Eq.~(\ref{eqn:Scaling_Factor_Omega_as_a_f(xo, to)}) to obtain
\begin{equation} \label{eqn:Omega-in-Polar-Coordinates}
     \Omega (\zeta_0,\sigma_0)= \frac{\zeta_0}{r_0}=\sqrt{\sech{2 \sigma_0}}\,.
\end{equation}
Note that $ \Omega (\zeta_0,\sigma_0) > 0$. The third projection corresponding to Eq. \ref{eqn:Mapping-points-p_2-to-p_3} now reads
\begin{eqnarray} \label{eqn:Polar-Mapping-points-p_2-to-p_3}
    &&p_2= (\Omega \zeta_0 \cosh{\sigma}_0,\Omega \zeta_0 \sinh{\sigma}_0,
           \zeta_0)\nonumber\\
  &\rightarrow &p_3  =(\Omega \zeta_0 \cosh{\sigma}_0,\Omega \zeta_0 \sinh{\sigma}_0, 0)\,. 
\end{eqnarray}

By suppressing the third coordinate corresponding to the $\zeta$-axis, the
combined point transformation from point $p_0$ to $\Bar{p}$ in
Eq.~(\ref{eqn:Complete-Projection-p0-to-pBar}) reads
\begin{eqnarray} \label{eqn:Complete-Mapping-points-p_0-to-p_bar}
    &&p_0= (\zeta_0 \cosh{\sigma}_0, \zeta_0 \sinh{\sigma}_0)\nonumber\\
  &\rightarrow &
           \Bar{p}  =(\Omega \zeta_0 \cosh{\sigma}_0,\Omega \zeta_0 \sinh{\sigma}_0)\,. 
\end{eqnarray}

On the light cones used in our first step of the projection, the hyperbolic
coordinate $\zeta_0$ is by definition the same as the new coordinate of our
auxiliary 3-dimensional spacetime, so far denoted by the same
letter. However, these two coordinates are independent of the light cones.
We will now begin to relabel the spacetime distance, $\zeta_0$, in the
$(x,t,\zeta)$ space, as $\xi_0$ in the $(x,t)$ and $(\Bar{x},\Bar{t})$
spaces. This notation allows us to consider vectors that are not necessarily
tangent to the light cones, and thereby to derive how an orthonormal frame,
which carries information about the metric, can be transformed by our
projections. In particular, we will be able to distinguish the tangent
vector, $\hat{\xi}$, in the 2-dimensional hyperplanes, along which
$\zeta_0=\xi_0$, from the direction in the third dimension, $\hat{\zeta}$,
depicted in Fig.~\ref{fig:conicproj}, along which the same distance $\zeta_0$
was being plotted thus far. We will thus rewrite
Eq.~(\ref{eqn:Complete-Mapping-points-p_0-to-p_bar}) as
\begin{eqnarray} \label{eqn:Complete-Mapping-points-p_0-to-p_bar-using-xi}
    &&p_0= (\xi_0 \cosh{\sigma}_0, \xi_0 \sinh{\sigma}_0) \nonumber\\
 &\rightarrow&
           \Bar{p}  =(\Omega \xi_0 \cosh{\sigma}_0,\Omega \xi_0 \sinh{\sigma}_0)\,. 
  \end{eqnarray}
Similarly, we obtain our mapping for a point in a timelike quadrant if we
simply flip $\cosh\sigma_0$ and $\sinh\sigma_0$:
\begin{eqnarray} \label{eqn:Complete-Mapping-points-p_0-to-p_bar-using-xiSPacelike}
  &&p_0= (\xi_0 \sinh{\sigma}_0, \xi_0 \cosh{\sigma}_0)\nonumber\\
  &\rightarrow& \Bar{p}=(\Omega \xi_0 \sinh{\sigma}_0,\Omega \xi_0 \cosh{\sigma}_0)
  \end{eqnarray}
  with the same $\Omega$. In both cases, the final point has Euclidean norm
  $\xi_0^2$ because $\sech(2\sigma_0)(\cosh^2\sigma_0+\sinh^2\sigma_0)=1$.

In Appendix~\ref{sec:RBS}, we show that this point projection is identical to a special case of the Renormalized Blended Spacetime (RBS) transformation proposed earlier in \cite{RBS}.

\subsection{Light cone} \label{sec:Lightcone}

So far, we considered points on hyperbolae in the Minkowski plane, assuming
non-zero Minkowski distance from the origin and excluding the light lines
defined by the points $p(x, \pm x)$ in the 2D Minkowski space. Our derived map
  \begin{equation} \label{m}
    m\colon p(x,t)\mapsto \bar{p} (\bar{x}, \bar{t})= \sqrt{\frac{|x^2-t^2|}{x^2+t^2}}(x,t)
  \end{equation}
  can be applied to the light lines as well, but it maps all its points to the
  origin of the Euclidean plane and is therefore not invertible. The Euclidean
  diagonals can then be interpreted as extended versions of the four points at
  infinity on the light lines.  As discussed in detail in \cite{RBS}
  $m^{-1}( \bar{p}(\bar{x}, \pm \bar{x}))={\rm sgn}(\bar{x})(\infty, \pm
  \infty)$, where $m^{-1}$ is the inverse of our mapping
  (\ref{m}). Alternatively, one may choose to apply the identity map on the
  light cone, resulting in an invertible but discontinuous complete
  mapping. In this paper, we are mainly interested in measure questions and
  therefore leave the question of how to map the light cone for further
  studies.

\section{Transformation of tangent  vectors}
\label{sec:Vectors}

\begin{figure*}[htbp]
    \centering
    \includegraphics[width=16cm]{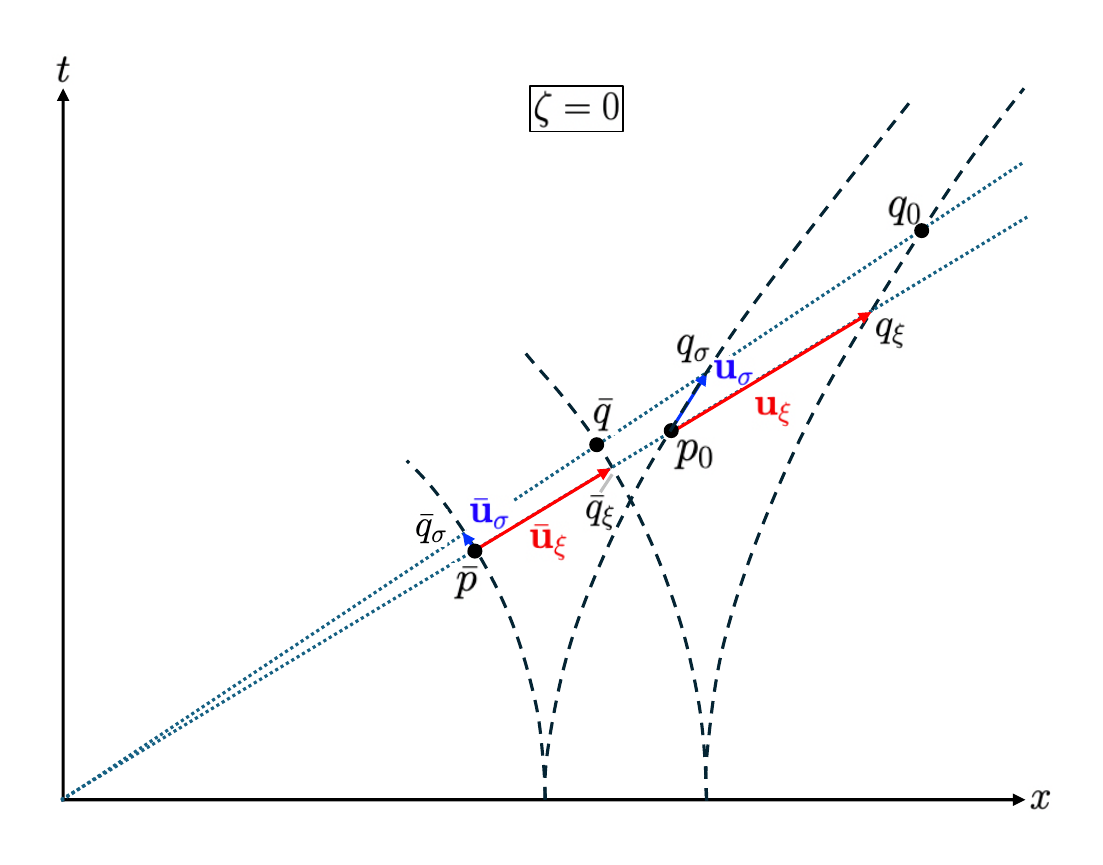}
    \caption{The tangent vectors, $\textbf{u}_\xi$,
      Eq.~(\ref{eqn:Tangent-vector-xi}), and $\textbf{u}_\sigma$,
      Eq.~(\ref{eqn:Tangent-vector-sigma}), at a point $p_0$ on the Minkowski
      plane, $\zeta=0$, are transformed by
      Eq.~(\ref{eqn:Complete-Mapping-points-p_0-to-p_bar-using-xi})
      respectively into the tangent vectors, $\Bar{\textbf{u}}_\xi$,
      Eq.~(\ref{eqn:Transformed-tangent-vector-xi}), and
      $\Bar{\textbf{u}}_\sigma$,
      Eq.~(\ref{eqn:Transformed-tangent-vector-sigma}), at the transformed
      point $\Bar{p}$ on a Euclidean plane that is coincident with the
      original Minkowski plane. The Lorentzian differential line element,
      ${\rm d}s=|{\rm d}\textbf{s}|=|\textbf{q}_0-\textbf{p}_0|$,
      Eq.~(\ref{eqn:Metric-line-element-MS}), transforms into the Euclidean
      differential line element,
      ${\rm d}\Bar{s}=|{\rm
        d}\Bar{\textbf{s}}|=|\Bar{\textbf{q}}-\Bar{\textbf{p}}|$,
      Eq.~(\ref{eqn:Line-element-dsBar-Euclidean}).}
    \label{fig:TangentVTrans}
\end{figure*}

We wish to transform the tangent space around every point $p_0$ in the
Minkowski $xt$-plane, to the Euclidean $\Bar{x}\Bar{t}$-plane. Consider an
arbitrary point $p_0$ in a spacelike quadrant of the Minkowski $xt$-plane whose
hyperbolic coordinates are given by
\begin{equation} \label{eqn:Polar-coordinates-of-Point-p}
    p_0=(\xi \cosh{\sigma}, \xi \sinh{\sigma})\,.
\end{equation}
This point is also represented by the vector, $\textbf{p}_0$, from the origin to the point $p_0$ in the Minkowski plane. Without loss of generality, we make a convenient choice of coordinate basis vectors that span the tangent space at the point $p_0$,
\begin{align} 
    \textbf{u}_{\xi} &= \, {\rm d} \xi \hat{\bm{\xi}}= \left ( \, {\rm d}\xi
                       \cosh{\sigma}, \, {\rm d}\xi \sinh{\sigma} \right ) \label{eqn:Tangent-vector-xi} \\
    \textbf{u}_{\sigma} &= \xi \, {\rm d} \sigma \hat{\bm{\sigma}} = 
\left ( \xi \, {\rm d}\sigma \sinh{\sigma}, \xi \, {\rm d}\sigma \cosh{\sigma}
                          \right ) \label{eqn:Tangent-vector-sigma} \,.
\end{align}
These tangent vectors are depicted in Fig.~\ref{fig:TangentVTrans}. The
norm-square of these basis vectors in the Minkowski plane are
\begin{widetext}
\begin{align} \label{eqn:Norms-of-tangent-basis-vectors}
    ||u_\xi||^2 &= \left ( \begin{array}{cc}
       \, {\rm d}\xi \cosh{\sigma}  &   \, {\rm d}\xi \sinh{\sigma} \\
    \end{array} \right ) \left ( \begin{array}{cc}
        1 & 0 \\
        0 & -1
    \end{array} \right ) \left ( \begin{array}{c}
        \, {\rm d}\xi \cosh{\sigma}  \\
         \, {\rm d}\xi \sinh{\sigma}
                                 \end{array} \right )= \, {\rm d}\xi^2 \\
    ||u_\sigma||^2 &= \left ( \begin{array}{cc}
       \xi \, {\rm d}\sigma \sinh{\sigma}  &   \xi \, {\rm d}\sigma \cosh{\sigma} \\
    \end{array} \right ) \left ( \begin{array}{cc}
        1 & 0 \\
        0 & -1
    \end{array} \right ) \left ( \begin{array}{c}
        \xi \, {\rm d}\sigma \sinh{\sigma}  \\
         \xi \, {\rm d}\sigma \cosh{\sigma}
                                 \end{array} \right )= - \xi^2 \, {\rm d}\sigma^2\,.
\end{align}
\end{widetext}

The metric tensor, $\eta_{\mu\nu}$, for the Minkowski plane is the $2 \times 2$ matrix above
with the diagonal elements of $+1$ and $-1$. The metric differential line
element is thus given by
\begin{equation} \label{eqn:Metric-line-element-MS}
    {\rm d}s^2 = ||u_\xi||^2+||u_\sigma||^2= \, {\rm d}\xi^2 - \xi^2 \, {\rm d}\sigma^2\,.
\end{equation}
The starting points for both of the infinitesimal basis vectors given by
Eqs.~(\ref{eqn:Tangent-vector-xi}) and (\ref{eqn:Tangent-vector-sigma}) are at
the point $p_0$ whose coordinates are given by
Eq.~(\ref{eqn:Polar-coordinates-of-Point-p}). Their ending points shown in Fig.~\ref{fig:TangentVTrans} are at
\begin{align} 
    q_\xi &= \left ((\xi+\, {\rm d}\xi) \cosh{\sigma}, (\xi+\, {\rm d}\xi)
            \sinh{\sigma} \right
            ) \label{eqn:End-Polar-Coordinates-for-u_xi}\\ 
    q_\sigma &= (\xi \cosh{\sigma}+\xi \, {\rm d}\sigma \sinh \sigma, \xi
               \sinh{\sigma}+\xi \, {\rm d}\sigma \cosh \sigma)\,. \label{eqn:End-Polar-Coordinates-for-u_sigma}
\end{align}

We now wish to perform point transformations
$p \rightarrow \Bar{p}=\Omega(p)p$,
$q_\xi \rightarrow \Bar{q}_\xi=\Omega(q_\xi)q_\xi$, and
$q_\sigma \rightarrow \Bar{q}_\sigma=\Omega(q_\sigma)q_\sigma$ as shown graphically in Fig.~\ref{fig:TangentVTrans}. The
transformation factors, $\Omega(p)$, $\Omega(q_\xi)$, and $\Omega(q_\sigma)$,
are determined by substituting the coordinates for the corresponding points
from Eqs.~(\ref{eqn:Polar-coordinates-of-Point-p}),
(\ref{eqn:End-Polar-Coordinates-for-u_xi}), and
(\ref{eqn:End-Polar-Coordinates-for-u_sigma}) into
Eq.~(\ref{eqn:Scaling_Factor_Omega_as_a_f(xo, to)}):
\begin{align} \label{eqn:Omega-factors-for-points-defining-tangent-vectors}
    \Omega(p) &= \sqrt{\sech{2 \sigma}} \\
    \Omega(q_\xi) &= \sqrt{\sech{2 \sigma}} \\
    \Omega(q_\sigma) &= \sqrt{\sech{2 \sigma}} \sqrt{\frac{1-\, {\rm
                       d}\sigma^2}{1+2\, {\rm d}\sigma \tanh{2\sigma}+\,{\rm d}\sigma^2}}
                       \nonumber \\ 
    & \approx \sqrt{\sech{2 \sigma}} \left ( 1- \, {\rm d}\sigma \tanh{2 \sigma} \right )\,.
\end{align}
Since ${\rm d}\sigma$ is infinitesimally small, in the last step we have ignored
terms of the order ${\rm d}\sigma^2$ in the previous expression, and Taylor-expanded
to first order in ${\rm d}\sigma$ about ${\rm d}\sigma=0$. Now we can proceed with the
following point transformations:
\begin{widetext}
\begin{align} 
\Bar{p} &= \left ( \Omega(p) \xi \cosh{\sigma}, \Omega(p) \xi \sinh{\sigma} \right ) = \left ( \sqrt{\sech{2 \sigma}} \xi \cosh{\sigma}, \sqrt{\sech{2 \sigma}} \xi \sinh{\sigma} \right )  \label{eqn:p-bar} \\
\Bar{q}_\xi &=\left (\Omega(q_\xi) (\xi+\, {\rm d}\xi) \cosh{\sigma},
              \Omega(q_\xi) (\xi+\, {\rm d}\xi) \sinh{\sigma} \right ) = \left
              (\sqrt{\sech{2 \sigma}} (\xi+\, {\rm d}\xi) \cosh{\sigma},
              \sqrt{\sech{2 \sigma}} (\xi+\, {\rm d}\xi) \sinh{\sigma} \right ) \label{eqn:q_xi-bar} \\
\Bar{q}_\sigma &= \left (\Omega(q)_\sigma (\xi \cosh{\sigma}+\xi \, {\rm
                 d}\sigma \sinh \sigma), \Omega(q)_\sigma (\xi
                 \sinh{\sigma}+\xi \, {\rm d}\sigma \cosh \sigma) \right ) \nonumber \\
& \approx \left ( \sqrt{\sech{2 \sigma}} \xi \cosh{\sigma} - \left (
                                                                                          \sqrt{\sech{2
                                                                                          \sigma}}
                                                                                          \right
                                                                                          )^3
                                                                                          \xi
                                                                                          \,
                                                                                          {\rm
                                                                                          d}\sigma
                                                                                          \sinh{\sigma}
                                                                                          ,
                                                                                          \sqrt{\sech{2
                                                                                          \sigma}}
                                                                                          \xi
                                                                                          \sinh{\sigma}
                                                                                          +
                                                                                          \left
                                                                                          (
                                                                                          \sqrt{\sech{2
                                                                                          \sigma}}
                                                                                          \right
                                                                                          )^3
                                                                                          \xi
                                                                                          \,
                                                                                          {\rm
                                                                                          d}\sigma
                                                                                          \cosh{\sigma} \right )\,. \label{eqn:q_sigma-bar}
\end{align}

The new basis vectors, $\Bar{\textbf{u}}_\xi$ and $\Bar{\textbf{u}}_\sigma$,
that span the transformed tangent space are given by the two vectors that
point from the point $\Bar{p}$ to the points $\Bar{q}_\xi$ and
$\Bar{q}_\sigma$, given by
\begin{align} 
    \Bar{\textbf{u}}_\xi &= \Bar{\textbf{q}}_\xi - \Bar{\textbf{p}} = \left
                           (\sqrt{\sech{2 \sigma}} \, {\rm d}\xi
                           \cosh{\sigma}, \sqrt{\sech{2 \sigma}} \, {\rm d}\xi \sinh{\sigma} \right ) \label{eqn:Transformed-tangent-vector-xi}\\
    \Bar{\textbf{u}}_\sigma &= \Bar{\textbf{q}}_\sigma - \Bar{\textbf{p}} =
                              \left ( - \left ( \sqrt{\sech{2 \sigma}} \right
                              )^3 \xi \, {\rm d}\sigma \sinh{\sigma} ,  \left
                              ( \sqrt{\sech{2 \sigma}} \right )^3 \xi \, {\rm d}\sigma \cosh{\sigma} \right )\,. \label{eqn:Transformed-tangent-vector-sigma}
\end{align}
They have been projected from a hyperbola to a circle, obtained as
cross-sections at constant $\zeta$ of light cones in two different
3-dimensional auxiliary Minkowski spacetime with line element
${\rm d}x^2-{\rm d}t^2-{\rm d}\zeta^2$, related to each other by flipping the
metric components $\eta_{xx}$ and $\eta_{\zeta\zeta}$ in the present case of a
spacelike quadrant. After this flip, a plane of constant $\zeta$ has a
negative Euclidean line element $-{\rm d}x^2-{\rm d}t^2$. We can therefore
compute the norms of our projected vectors using the negative identity matrix
as the tangent-space metric.
Their negative-Euclidean norm-squares of our basis vectors  then given by
\begin{align} \label{eqn:Norms-of-transformed-tangent-vectors}
    ||\Bar{\textbf{u}}_\xi||^2 &=  \left ( \begin{array}{cc}
       \sqrt{\sech{2 \sigma}} \, {\rm d}\xi \cosh{\sigma}  &   \sqrt{\sech{2
                                                             \sigma}} \, {\rm d}\xi \sinh{\sigma} \\
    \end{array} \right ) \left ( \begin{array}{cc}
        -1 & 0 \\
        0 & -1
    \end{array} \right ) \left ( \begin{array}{c}
       \sqrt{\sech{2 \sigma}} \, {\rm d}\xi \cosh{\sigma}  \\
         \sqrt{\sech{2 \sigma}} \, {\rm d}\xi \sinh{\sigma}
    \end{array} \right ) = \, -{\rm d}\xi^2 \\
    || \Bar{\textbf{u}}_\sigma ||^2 &= \left ( \begin{array}{cc}
       - \left ( \sqrt{\sech{2 \sigma}} \right )^3 \xi \, {\rm d}\sigma
                                                 \sinh{\sigma}   &   \left (
                                                                   \sqrt{\sech{2
                                                                   \sigma}}
                                                                   \right )^3
                                                                   \xi \, {\rm
                                                                   d}\sigma \cosh{\sigma} \\
    \end{array} \right ) \left ( \begin{array}{cc}
        -1 & 0 \\
        0 & -1
    \end{array} \right ) \left ( \begin{array}{c}
       - \left ( \sqrt{\sech{2 \sigma}} \right )^3 \xi \, {\rm d}\sigma \sinh{\sigma}   \\
         \left ( \sqrt{\sech{2 \sigma}} \right )^3 \xi \, {\rm d}\sigma \cosh{\sigma}
    \end{array} \right ) \nonumber \\
    &= -\sech^2(2 \sigma) \xi^2 \, {\rm d}\sigma^2\,.
\end{align}
\end{widetext}

Up to the sign, the transformed basis vector $\Bar{\textbf{u}}_\xi$ has the same norm as that
of the original basis vector $\textbf{u}_\xi$ before the
transformation. However, the transformed basis vector
$\Bar{\textbf{u}}_\sigma$ has a norm that is scaled by a factor of
$\sech{2 \sigma}$ relative to the original basis vector, $\textbf{u}_\sigma$,
before the transformation. The metric differential line element
suggested by the transformation is
\begin{equation} \label{eqn:Line-element-dsBar-Euclidean}
    \, {\rm d}\Bar{s}^2 = ||\Bar{\textbf{u}}_\xi||^2+||\Bar{\textbf{u}}_\sigma||^2= \, -{\rm d}\xi^2 - \sech^2(2 \sigma) \xi^2 \, {\rm d}\sigma^2 
\end{equation}
and can be transformed into a positive-definite expression by a conformal sign
change, multiplying all terms by $-1$.
If we define ${\rm d}\phi=\sech(2\sigma){\rm d}\sigma$, or
\begin{equation} \label{phisigma}
  \phi= \frac{1}{2}\arctan(\sinh(2\sigma))\,,
\end{equation}
we
obtain the Euclidean line element in its standard polar form. The range of $\phi$ is
$\pi/2$, which is correct for a quarter circle obtained from a single timelike
or spacelike sector in the original Minkowksi spacetime. As suggested by the
central projection of a timelike or spacelike sector in Minkowski spacetime to
a quadrant of the Euclidean plane, the non-compact hyperbolas along which
$\sigma$ changes at constant $\xi$ are mapped to compact quarter-circles
with a finite $\phi$-range. The factor of $\sech(2\sigma)$ accomplishes this
transformation with the correct quantitative relationships between the ranges.

If we repeat the construction for a timelike quadrant, a point has hyperbolic
coordinates $(\xi\sinh\sigma,\xi\cosh\sigma)$ and implies the original Minkowski
line element $-{\rm d}\xi^2+\xi^2{\rm d}\sigma^2$. The 3-dimensional auxiliary
spacetime in this case has the line element ${\rm d}x^2-{\rm d}t^2+{\rm
  d}\zeta^2$, and the transition to Euclidean signature on constant-$\zeta$
planes is obtained by flipping $\eta_{tt}$ and $\eta_{\zeta\zeta}$. The
resulting line element is positive-Euclidean, ${\rm d}x^2+{\rm d}t^2$, such
that we obtain positive Euclidean norm-squares for the projected basis vectors.
The line element is therefore  mapped to
${\rm d}\xi^2+\sech^2(2\sigma)\xi^2{\rm d}\sigma^2$ without the need of a
conformal sign change. It can be mapped to the standard Euclidean line element
by applying the same coordinate transformation from $\sigma$ to $\phi$
as used in spacelike quadrants. In all cases, we therefore arrive at the line
element ${\rm d}\xi^2+\xi^2{\rm d}\phi^2={\rm d}x^2+{\rm d}y^2$ with
$x=\xi\cos\phi$ and $y=\xi\sin\phi$. The change of signature happens in our
{\em second projection} where we map between two different cones, or their
cross-sections shown in Fig.~\ref{fig:TangentVTrans}, in our
3-dimensional auxiliary spacetime. 

\subsection{Summary}
\label{sec:Mapping}

The combination of our individual transformation steps has resulted in a
mapping from 2-dimensional Minkowski spacetime with line element
${\rm d}x^2-{\rm d}t^2$ to 2-dimensional Euclidean space with line element
${\rm d}x^2+{\rm d}y^2$. We performed the mapping separately in the four
quadrants of Minkowski spacetime, given by two spacelike and two timelike
ones. Working with projections, it was convenient to express Minkowski
spacetime in a hyperbolic version of polar coordinates, in which the line
element takes the form $\pm{\rm d}\xi^2\mp\xi^2{\rm d}\sigma^2$. A radial
projection that preserves each spacelike and timelike quadrant can then be
performed in a fixed coordinate direction along $\xi$.

This result provides our basic ingredient for transformations of integrals as
they may appear in different ways in quantum field theories. There are two
implications. 

First, the radial projection between two different cones in the
auxiliary 3-dimensional spacetime not only implements the crucial sign change,
it also maps a non-compact hyperbola to a compact circle segment. Accordingly,
the line elements in polar coordinates are related by the Minkowskian $\pm{\rm
  d}\xi^2\mp\xi^2{\rm d}\sigma^2$ being mapped to $\mp({\rm d}\xi^2 +
\sech^2(2 \sigma) \xi^2 \, {\rm d}\sigma^2)$ with a rescaling factor of
$\sech^2(2\sigma)$ in the angular part. We already showed that this is
precisely the factor required for the coordinate transformation
(\ref{phisigma}) to result in the standard Euclidean angle $\phi$ with the
correct range for a quarter-circle in each quadrant. In
Section~\ref{sec:Integrals}, we will demonstrate that this factor also
correctly regularizes divergent QFT integrals and implies the same result as a
Wick rotation, but without complex extensions of coordinates.

The second implication is important for action integrals $S$ and their
interpretation as exponents in a path integral or a statistical sum. A
standard Wick rotation is motivated by the desire to turn a complex amplitude
$\exp(iS/\hbar)$ in a path integral into a real-valued statistical factor $\exp(-S_{\rm
  E}/\hbar)$ with the Euclidean action $S_{\rm E}$ corresponding to $S$, in
which a field theory is formulated on Euclidean space instead of Minkowski
spacetime. In Section~\ref{s:App}, we will demonstrate in several examples
that the same transition is accomplished by an application of our new mapping,
based on the following ingredients extracted from our explicit construction.

Through the steps of our transformation, metric components in Minkowski
spacetime are directly mapped to Euclidean components, as justified by our
explicit transformations between points and directions.  For compact notation,
we write the line elements as
${\rm d}s_{\rm Minkowski}^2=\eta_{\mu\nu}{\rm d}x^{\mu}{\rm d}x^{\nu}$ and
${\rm d}s_{\rm Euclid}^2=\delta_{ij}{\rm d}y^i{\rm d}y^j$, respectively, with
collective coordinates $x^{\mu}$ and $y^i$, and use the convention that
repeated indices are summed over their respective ranges. The metric
transformation then takes the form
\begin{equation} \label{eqn:MetricTransform}
  \eta_{\mu\nu} \mapsto \delta_{ij}\,,
\end{equation}
A given metric-dependent Lorentzian action can be transformed by direct
substitution of (\ref{eqn:MetricTransform}) if we only make sure that the
action explicitly contains metric factors required on a generic spacetime,
such as $\sqrt{-\det(\eta_{\mu\nu})}$ in the volume element, even if they
equal one on Minkowski spacetime.  The transformation
(\ref{eqn:MetricTransform}) is then applied directly while leaving all other
fields and coefficients unchanged. In particular, the determinant
$\det(\eta_{\mu\nu})=-1$ is mapped to $\det (\delta_{ij})=1$. By this
replacement, the volume element in a spacetime integral, with a factor of
$\sqrt{-\det(\eta_{\mu\nu})}=1$, is mapped to the imaginary unit
\begin{equation} \label{eqn:deltai}
  \sqrt{-\det(\delta_{ij})}=-i\,.
\end{equation}
(We will see in Section~\ref{s:App} that the negative root is suitable for action principles.)
This factor of $i$ can be used to replace the
use of imaginary time in a Wick rotation. We will describe such applications
in more detail in Section~\ref{s:App}, and also comment on actions that cannot
be written directly in terms of a metric, in particular on fermionic theories
that require a tetrad.

\section{Integrals}
\label{sec:Integrals}

We now demonstrate
that the our transformation correctly relates Lorentzian and Euclidean
integrals as they commonly appear in calculations for quantum field theories
\cite{QFT}. These integrals, used for instance for determining loop
corrections to propagators, are of the form
\begin{equation} \label{eqn:int}
    \int \text{d}^dq \; f (q^2)
\end{equation}
where $d$ is the dimension of spacetime with a Lorentzian metric
signature, $q$ is a spacetime coordinate vector, and $f(q^2)$ is a function depending
only on the coordinate distance from the origin,  given by 
\begin{equation}
    q^2 = -q_0^2+q_1^2+\cdots+q_{d-1}^2\,.
\end{equation}
If the integrand has poles, as in loop corrections, the
  expression $f(q^2)$ has to be adjusted in order to move poles away from the
  integration contour, as in $f(q^2+i\epsilon)$. For now, we focus on a more
  basic feature and compare the integration measures or area elements on the
  Minkowskian and Euclidean sides of our mapping, assuming that $f(q^2)$ does
  not have poles. We will then see that the case of $f(q^2)$ with poles can
  also be discussed successfully by our mapping.

The standard technique for computing an integral of this form is a Wick
rotation that transforms $q$ by rotating its time-component by $\pi/2$ in the
complex plane, reversing the negative sign in $q^2$ and resulting in a
$d$-dimensional Euclidean integral (multiplied by $i$). In some cases, the
results obtained in this way can be justified by an application of Cauchy's
theorem, using a complex extension of the integrand to a holomorphic function
and rotating integration over the real axis to integration over the imaginary
axis based on the $\epsilon$-prescription. However, Wick
rotation is also applied to non-convergent integrals, which are turned into
convergent integrals by this procedure. In general, therefore, Wick rotation
is a combination of Cauchy's theorem with regularization steps that are often
spelled out incompletely, producing a finite value without a clear
relationship to the original integral.  Moreover, this procedure does not
illuminate the geometrical significance of Euclidean geometry.  Notably, in
contrast to Lorentz boosts, for instance, there is no physical spacetime
transformation that produces a Wick rotation. Our geometrical construction can
be used to shed light on these questions.

For now, we will set $d=2$, with $q=(q_0,q_1)=(t,x)$. Later we will address generalization to higher dimensions. Our integral now takes the form
\begin{equation}
   I = \int ^\infty _{-\infty}  \int_{-\infty}^\infty \text{d}x \text{d}t \, f(x^2-t^2).
 \end{equation}
As mentioned, we first assume that the function $f$ does not
   have poles and focus on measure questions of our 2-dimensional
   integrations. To this end,
we convert to hyperbolic coordinates $(\xi,\sigma)$ such that
$x=\pm\xi\cosh\sigma$ and $t=\pm\xi\sinh\sigma$ for the two spacelike
quadrants, and $x=\pm\xi\sinh\sigma$ and $t=\pm\xi\cosh\sigma$ for the two
timelike quadrants. In each quadrant, ${\rm d}x{\rm d}t=\xi{\rm d}\xi{\rm
  d}\sigma$, and we have $x^2-t^2=\xi^2$ in spacelike quadrants while
$x^2-t^2=-\xi^2$ in timelike quadrants. Since $\xi$ runs from zero to $\infty$
and $\sigma$ from $-\infty$ to $\infty$ in each quadrant, the full spacetime
integral is expressed as 
\begin{equation} \label{eq:I-sum}
    I = 2 \int_{-\infty}^\infty \text{d}\sigma \int_0^\infty \text{d}\xi \,
    \xi \, \left( f(\xi^2) - f(-\xi^2) \right)
  \end{equation}
  in hyperbolic coordinates.

\subsection{Angular Part}

Since the whole integral shares a single angular part that is not dependent on
the function of integration, it can be computed independently. However, as
written so far, the angular part in  Eq.~(\ref{eq:I-sum}) diverges. This
divergence is directly related to the non-compact directions coordinatized by
$\sigma$ along which the integrand is constant. Similarly, the Cartesian
expression has a constant integrand along non-compact curves along which
$x^2-t^2={\rm constant}$. These directions are not suppressed by measure
factors in the original Minkowski geometry.

However, our transformed geometry contains an additional factor of
$\sech(2\sigma)$ in the metric which, as we already showed, maps non-compact
hyperbolae to compact quarter circles. An integration of a constant function
over this compact range is guaranteed to be finite, but we should also make
sure that the precise value is correct, for instance by comparing it with the
result of a Wick rotation traditionally used in QFT calculations. We should
then include any additional factor obtained from the metric transformation
(\ref{eqn:MetricTransform}), which in this case implies an additional factor
of $-i$ because the Minkowskian area element $\sqrt{-\det(\eta_{\mu\nu})}$ in
Cartesian coordinates is mapped to $\sqrt{-\det(\delta_{ij})}=-i$ if we use
the sign convention (\ref{eqn:deltai}). Since we have written the integral in
polar coordinates $(\sigma,\xi)$, the measure contains an additional factor of
the Jacobian of the transformation, which is equal to $\xi$ in the Minkowskian
hyperbolic system, and equal to $\sech(2\sigma) \xi$ in the Euclidean mapping,
where the line element is given by (\ref{eqn:Line-element-dsBar-Euclidean}).

Applying this transformation as an
explicit regularization step, the 
integral takes the form
\begin{equation}\label{eq}
    I = -2i \int_{-\infty}^\infty \frac{\text{d}\sigma}{\cosh{2\sigma}} \int_0^\infty \text{d}\xi \, \xi \, \left( f(\xi^2) - f(-\xi^2) \right)\,.
\end{equation}
The angular integral can be easily computed:
\begin{eqnarray}
  \int_{-\infty}^\infty \frac{\text{d}\sigma}{\cosh{2\sigma}} &=&
  \frac{1}{2}  \int_{-\infty}^{\infty} \frac{{\rm
                                                                  d}\sinh(2\sigma)}{1+\sinh^2(2\sigma)}\nonumber\\
  &=&\frac{1}{2}\arctan(\sinh(2\sigma))|_{-\infty}^{\infty}=
  \frac{\pi}{2}\,.
\end{eqnarray}
The result,
\begin{equation} \label{eqn:intAfterAngle}
     I = -\pi i \int_0^\infty \text{d}\xi \, \xi \, \left( f(\xi^2) - f(-\xi^2) \right)\,,
\end{equation}
is equivalent to the standard integral obtained via Wick rotation, from which
it differs only by the choice of the sign in (\ref{eqn:deltai}).

Our derivation makes clear how the Euclidean integral differs from the
Lorentzian one, owing to an additional factor of $-i\sech(2\sigma)$ in the
measure. As an alternative regularization, we could use the Lorentzian
$\sigma$-independent area element but choose a different topology such that
hyperbolas with the infinite extension of $\sigma$ are compact. Specifically,
the Bohr compactification of the real line \cite{Bohr} fulfills this purpose,
according to which the original integration
$\int_{-\infty}^{\infty}{\rm d}\sigma$ is replaced by
\begin{equation}
  \lim_{N\to\infty}\frac{1}{2N}
  \int_{-\frac{1}{2}N\pi}^{\frac{1}{2}N\pi}{\rm d}\sigma\,.
\end{equation}
The integration of a $\sigma$-independent function with this measure produces the same value of
$\pi/2$ as the Euclidean integration. In this case, however, we have to adjust
the integration limits $\pm\frac{1}{2}N\pi$ by hand in order to produce this
value, while the correct result was obtained automatically by the canonical
projections to Euclidean geometry. Moreover, we do not obtain a factor of $i$
if we merely compactify the original integration range. Our full geometrical
transformation is therefore important for applications to integrals that
appear in quantum field theory.

\subsection{Radial Part}

Our discussion of the integration measure shows a correct
  mapping from Minkowskian to Euclidean results. For a more detailed
  comparison with QFT results, we should also consider integrands with
  poles. We now show that our mapping can directly be applied to such relevant
  functions as well.

In the case of loop corrections, the function $f$ is of the form
\begin{equation}
    f(q^2)=\frac{q^{2a}}{(q^2+D)^b}
\end{equation}
where $D\not=0$ is a real constant, $a$ and $b$ are nonnegative integers, and
$b>a+1$ (to ensure convergence of this integral).
The Minkowski version of this function has poles because $q^2$
  is not positive definite in this case. A unique value is obtained for a
  suitable choice of an $\epsilon$-prescription.

Our task is now to evaluate the ``radial'' integrals $I_{SL}$ and $I_{TL}$,
defined by
\begin{equation}
    I_{SL}:=\int_0^\infty \frac{\text{d}\bar{\xi}\,\bar{\xi}^{2a+1}}{\left(\bar{\xi}^2+D\right)^b}
\end{equation}
and
\begin{equation}
    I_{TL}:=\int_0^\infty \frac{\text{d}\bar{\xi}\,\bar{\xi}\,(-\bar{\xi}^2)^a}{\left(-\bar{\xi}^2+D\right)^b}\,.
  \end{equation}
Depending on the sign of $D$, the integrand of either $I_{SL}$ or $I_{TL}$ has
  a pole, and therefore the integral depends on its treatment. Choosing the
  principal value, the results for the two integrals differ by only a sign:
\begin{equation}
    I_{SL}= \frac{\Gamma(a+1)\Gamma(b-a-1)}{2 D^{b-a-1} \Gamma(b)}
\end{equation}
and
\begin{equation}
    I_{TL}= -\frac{\Gamma(a+1)\Gamma(b-a-1)}{2 D^{b-a-1} \Gamma(b)}.
\end{equation}

Thus, the whole integral evaluates to
\begin{equation}
    I=-\pi i \frac{\Gamma(a+1)\Gamma(b-a-1)}{D^{b-a-1} \Gamma(b)}
\end{equation}
which, up to the sign choice, is precisely the result determined by the
standard method \cite{QFT}. The sign difference means that complex amplitudes
in which such integrals appear
are replaced with their complex conjugates, which does not imply physical
differences in probabilities.

\subsection{Lightcone singularities}

For $D=0$, our integrals do not converge for any values of $a$
  and $b$. This outcome is a result of the limiting behavior of our mapping
  which maps the entire lightcone, and therefore any lightlike pole, to the
  origin of the Euclidean plane. Any lightlike pole then appears at the
  boundary of the $\bar{\xi}$-integrations and principal values are no longer
  available. A modification of our mapping on the lightcone might possibly be
  used to achieve finite integrals, but it would require detailed
  considerations of integrands that we leave for future work.

\section{Applications in field theory and gravity}
\label{s:App}

So far,  our transformation  has been  formulated for  a 2-dimensional Minkowski
spacetime. If we  use polar  coordinates in  the spatial  part of  Minkowski
spacetime,  we  can  directly  extend   the  2-dimensional  mapping  to  four
dimensions, where  $(r,t)$ now play  the role of  the previous $(x,t)$  in our
derivation. Since the final transformation  is specified for metric components
and affects only $\eta_{tt}$, we do not need to write field theories in
spatial polar coordinates.

The resulting metric transformation (\ref{eqn:MetricTransform}), now in four
dimensions, can directly be applied to standard scalar or Yang-Mills actions
in Minkowski spacetime, which only require the metric tensor. We will
briefly make these examples more explicit, and then turn to more involved
cases of fermions, which require a tetrad and gamma matrices, as well
as field theories at constant chemical potential. We will also give an outlook
on possible extensions to quantum fields on curved spacetime as well as
gravity itself in which case the original metric is no longer Minkowskian.

\subsection{Scalar and Yang--Mills theories}

The scalar action
\begin{equation}
  S=-\int{\rm d}^4x \sqrt{-\det\eta}
  \left(\frac{1}{2}\eta^{\mu\nu}(\partial_{\mu}\phi)(\partial_{\nu}\phi)+V(\phi)\right)
\end{equation}
is transformed to
\begin{equation}
  iS_{\rm E}= i\int{\rm d}^4x
  \sqrt{\det\delta}\left(\frac{1}{2}\delta^{ij}(\partial_i\phi)(\partial_j\phi)+V(\phi)\right)
\end{equation}
following the sign convention (\ref{eqn:deltai}). The complex amplitude
$\exp(iS/\hbar)$ in a path integral is then turned into the real statistical
weight $\exp(-S_{\rm E}/\hbar)$.

The Yang--Mills action
\begin{equation}
  S=-\frac{1}{4}\int{\rm d}^4x \sqrt{-\det \eta}\;
  \eta^{\mu\rho}\eta^{\nu\sigma} F^a_{\mu\nu}F^a_{\rho\sigma}\,,
\end{equation}
with $F^a_{\mu\nu}=\partial_{\mu}A^a_{\nu}-\partial_{\nu}A^a_{\mu}+g
f^{abc}A_{\mu}^bA_{\nu}^c$ in terms of the coupling constant $g$ and structure
constants $f^{abc}$,
is similarly transformed to
\begin{equation}
  iS_{\rm E}=\frac{1}{4}i\int{\rm d}^4x \sqrt{\det\delta}\;
  \delta^{ik}\delta^{jl} F^a_{ij}F^a_{kl}\,.
\end{equation}
This expression with $f^{abc}=0$ can also be used for the electromagnetic field.

Perturbation theory applied to these Euclideanized actions
  then implies results equivalent to what is usually obtained after a Wick rotation.

\subsection{Fermions}

The Wick rotation of fermionic theories is not as straightforward as that of
bosonic theories, and different versions have been proposed as discussed for
instance in \cite{FermionWick}. The Lorentzian action of a Dirac spinor $\psi$
is given by
\begin{equation} \label{Sfermion}
  S=-\int{\rm d}^4x \sqrt{-\det\eta} \; \psi^{\dagger}
  i\gamma^0(\gamma^{I}e_I^{\mu}\partial_{\mu}+m)\psi
\end{equation}
with Dirac's gamma matrices obeying the anti-commutators
$\{\gamma^I,\gamma^J\}=2\eta^{IJ}$. With our sign choice, $\gamma^0$ is
anti-Hermitian, while $\gamma^1$, $\gamma^2$ and $\gamma^4$ are Hermitian. The
tetrad components $e_I^{\mu}$, such that
$\eta^{IJ}e_I^{\mu}e_J^{\nu}=\eta^{\mu\nu}$, relate the $\gamma^I$ to
tangent-space directions. The internal indices $I$ and $J$ are contracted with
a constant Minkowski metric $\eta^{IJ}$ formally unrelated to
$\eta^{\mu\nu}$. We will apply our geometrical transformation of metrics only
to the tangent-space metric $\eta_{\mu\nu}$ but not directly to the internal
metric $\eta^{IJ}$ because the latter does not provide a geometry on the
spacetime manifold. Nevertheless, a substitution of the internal Minkowski
$\eta^{IJ}$ by a Euclidean $\delta^{ab}$ will automatically follow from the
combined transformation of all terms in the fermion action.

The tetrad constitutes a new ingredient which is related to the metric and
should therefore participate in our transformation. Heuristically, it may be
considered a square root of the metric, such that sign changes in metric
components are related to factors of $i$ in the tertad. Similar
transformations in the context of Wick rotations have been used in
\cite{EuclideanSpinor}. At this point, complex numbers appear in the
transformation, but only when applied to fermions which are already formulated
with complex spinors in the Lorentzian theory. In detail, when we perform the
central step of our transformation, flipping $g_{tt}$ or $g_{xx}$ with
$g_{\zeta\zeta}$ in the auxiliary spacetime, followed by a conformal sign
change of the Euclidean metric in the spacelike case, we replace the time
component $e_0^{\mu}$ with $if_4^i$ as the fourth components of a Euclidean
tetrad $f_a^i$. The remaining tetrad vectors are directly related by mapping
the spacelike $(e_1^{\mu},e_2^{\mu},e_3^{\mu})$ to $(f_1^i,f_2^i,f_3^i)$.  In
this way, the factors of $i$ in the transformed $e_0^{\mu}$ will make sure
that the internal $\eta^{IJ}$ is replaced by the Euclidean $\delta^{ab}$:
\begin{eqnarray}
  &&\eta^{\mu\nu}=\eta^{IJ} e_I^{\mu} e_J^{\nu}=
  -e_0^{\mu}e_0^{\nu}+e_1^{\mu}e_1^{\nu}+
     e_2^{\mu}e_2^{\nu}+e_3^{\mu}e_3^{\nu} \nonumber\\
  &\mapsto& -i^2
  f_4^if_4^j+f_1^if_1^j+f_2^if_2^j+f_3^if_3^j= \delta^{ab}f_a^if_b^j =\delta^{ij}\,.
\end{eqnarray}
The internal Euclidean metric therefore appears automatically, without
directly mapping its components by conic projections. Moreover, we do not need
to introduce a separate tetrad for the internal metric.

The internal Minkowski metric also appears in the defining relation
$\{\gamma^I,\gamma^J\}=2\eta^{IJ}$ of the gamma matrices. Here, this metric is
automatically turned into the internal Euclidean metric $\delta^{ab}$ if we
define $i\gamma^0=\gamma^4$, which is also a standard step in a Wick rotation
of fermions \cite{FermionWick}. In this way, the anticommutators
$\{\gamma^I,\gamma^J\}=2\eta^{IJ}$ are mapped to
$\{\gamma^a,\gamma^b\}=2\delta^{ab}$. There is another $\gamma^0$ in the
derivative operator $\gamma^I e_I^{\mu}\partial_{\mu}$, which together with
the complex tetrad transformation is mapped according to
\begin{eqnarray}
&&  \gamma^Ie_I^{\mu}\partial_{\mu}=(\gamma^0e_0^{\mu}+\gamma^1e_1^{\mu}+
   \gamma^2e_2^{\mu}+\gamma^3e_3^{\mu})\partial_{\mu}\nonumber\\
  &\mapsto& (\gamma^0 (if_4^{i})+\gamma^1f_1^{i}+
   \gamma^2f_2^{i}+\gamma^3f_3^{i})\partial_{i}=\gamma^af_a^i\partial_i\,. \label{gammaef}
\end{eqnarray}

We are now ready to combine the individual contributions to the action
(\ref{Sfermion}). As in the scalar and Yang--Mills cases, the volume element
provides a factor of $\sqrt{-1}=-i$ which will appear as a factor of a
Euclidean fermion action in terms of Euclidean gamma matrices and a Euclidean
tetrad.  The resulting transformation to the (non-Hermitian) Euclidean action
\begin{equation} \label{SEFermion}
  S_{\rm E}=\int{\rm d}^4x \; \psi^{\dagger}
  \gamma^4(\gamma^{a}f_a^{i}\partial_{i}+m)\psi
\end{equation}
is related to a Wick
rotation discussed in \cite{EuclideanSpinor}. Unlike bosonic theories,
fermionic theories are not transformed uniquely by our mapping (just as for
Wick rotations) because changes of the tetrad components may be accompanied by
spinor transformations or redefinitions of the gamma matrices, as for instance in
\cite{EuclideanQED,FermionWick}. As in these previous proposals, a Hermitian
action would be obtained if the transformation of $e_I^{\mu}$ is
accompanied by a suitable transformation of the spinors, such that
$\gamma^4f_4^i$ in (\ref{SEFermion}) is replaced by $\gamma_{\rm E}^4f_4^i$ with
$\gamma_{\rm E}^4=\gamma^5=i\gamma^0\gamma^1\gamma^2\gamma^3$ while
$\gamma_{\rm E}^1=\gamma^1$, $\gamma_{\rm E}^2=\gamma^2$ and $\gamma_{\rm
  E}^3=\gamma^3$ remain unchanged. The explicit $\gamma^4=:\gamma_{\rm E}^5$
is then Hermitian and anticommutes with all four $\gamma_{\rm E}^i$, such that
\begin{equation} \label{SEFermionE}
  S_{\rm E}=\int{\rm d}^4x \; \bar{\psi}_{\rm E}(\gamma_{\rm E}^{a}f_a^{i}\partial_{i}+m)\psi
\end{equation}
with $\bar{\psi}_{\rm E}=\psi^{\dagger}\gamma_{\rm E}^5$ is real.

An interesting open question is whether the spinor transformation constructed
in \cite{FermionWick} has a geometrical analog related to our mapping of
spacetime geometries.

\subsection{Field theories at constant chemical potential}

Many-body states with varying particle numbers require action principles at
a chemical potential $\mu$, together with a term $-\mu\rho$ with the particle
density $\rho$ added to the integrand. In relativistic theories, the density
is the time component of a current $j^{\mu}$, which for a charge-$q$, and
complex-valued scalar field $\phi$
is given by
\begin{equation}
  j^{\mu}=i\hbar q \eta^{\mu\nu}
  (\phi^*\partial_{\nu}\phi-\phi\partial_{\nu}\phi^*)\,.
\end{equation}
The current may then appear in coupling terms of the form $A_{\mu}j^{\mu}$,
such that $A_0$ plays the role of a chemical potential $\mu$.

A Wick rotation would turn the time component of this expression into a
complex number because it is of first order in time derivatives. With our
transformation, we only change the sign of the time component through
$\eta^{00}$, turning the negative action contribution $-\mu j^0$ into the
Euclidean contribution $\mu j^0$.

For fermionic theories, the current is given by
$j^{\mu}= \bar{\psi}\gamma^Ie^{\mu}_I\psi$ with
$\bar{\psi}=\psi^{\dagger}\gamma^0$. The time component $j^0$ contains two
factors of $\gamma^0$, one from $\bar{\psi}$ and one from $\gamma^Ie_I^{\mu}$,
but they play different roles in our transformation because only one of them
is multiplied by the tetrad $e^{\mu}_0$. According to our fermionic
application, the term $\gamma^Ie^{\mu}_I$ is turned into the Euclidean
$\gamma^af_a^i$ as in (\ref{gammaef}). The $\gamma^0$ in $\bar{\psi}$,
however, is not accompanied by tetrad components and should directly be
replaced with $\gamma^0=-i\gamma^4$ in order to obtain Euclidean gamma
matrices. The current is then transformed into $-i\psi^{\dagger}\gamma^4
\gamma^af_a^i\psi$, which is not Hermitian. However, if the tretrad
transformation is accompanied by a transformation of spinors, as in
\cite{FermionWick}, that turns $\gamma^4f_4^i$ into $\gamma^5f_4^i=\gamma_{\rm
  E}^4f_4^i$ and identifies
$\gamma^4=\gamma_{\rm E}^5$, the transformed
current $j_{\rm E}^{\mu}= -i\bar{\psi}_{\rm E}
\gamma_{\rm E}^af_a^i\psi$ is Hermitian, and a real contribution $\mu j_{\rm E}^4$ is obtained at
constant chemical potential.

\subsection{Curved backgrounds and gravity}

Field theories on curved background spacetimes as well as gravity require
non-Minkowskian metrics $g_{\mu\nu}$ of Lorentzian signature. Generically, the
metric components may depend on all coordinates, including time, such that
they would become complex if a Wick rotation or possible generalizations such
as \cite{NewWick} are applied.

Since our procedure does not use complex numbers in the spacetime
transformation, it avoids complex metric components. Moreover, its geometrical
nature implies that the main ingredients, at least in two dimensions, can be
generalized to such theories.  We used only constructions such as curves of
equal distance from the origin or straight lines and planes for projections
that can be generalized to geodesic properties in a curved geometry. For the
extension to four dimensions we used a shortcut of introducing spatial polar
coordinates, which in general are not available unless the metric is
spherically symmetric. The general case would therefore require a
4-dimensional construction, which is harder to visualize and therefore
postponed to later work.

Continuing for now with the case of two spacetime dimensions, any curved
spacetime is conformally flat, such that we can find a spacetime function
$\lambda(t,x)$ that brings the line element to the form
\begin{equation}
  g_{\mu\nu}{\rm d}x^{\mu}{\rm d}x^{\nu}=\lambda(x,t)^2({\rm
    d}x^2-{\rm d}t^2)\,.
\end{equation}
Our procedure then goes through almost unchanged: We merely have to factor out
$\lambda^2$ and apply flipping $g_{tt}$ or $g_{xx}$ with the auxiliary
$g_{\zeta\zeta}$ only in the flat factor, using the 3-dimensional metrics
$\lambda^2({\rm d}x^2-{\rm d}t^2\pm {\rm d}\zeta^2)$ for the auxiliary spacetime. 

This procedure does not have a direct extension to more than two dimensions
because generic line elements then are not conformally flat. A suitable
auxiliary spacetime of five or more dimensions may then be constructed in
different ways. For instance, for timelike regions around a given point, we
may use Gaussian normal coordinates that bring the line element to the form
${\rm d}s^2=-{\rm d}t^2+{\rm d}\sigma^2$ with a positive curved spatial line
element ${\rm d}\sigma^2$, and define a 5-dimensional auxiliary spacetime as
${\rm d}s^2+{\rm d}\zeta^2$ with a single auxiliary coordinate $\zeta$. For
spacelike regions, we may use a 7-dimensional auxiliary spacetime in which all
three spatial coordinates are treated on an equal footing, and therefore
accompanied by three auxiliary dimensions. We then define the line element
${\rm d}s^2-{\rm d}\sigma_{\zeta}^2$ where ${\rm
  d}\sigma_{\zeta}^2$ is obtained from ${\rm d}\sigma^2$ by replacing the
spatial coordinates $(x_1,x_2,x_3)$ (which in general are no longer Cartesian)
with a triplet of auxiliary coordinates, $(\zeta_1,\zeta_2,\zeta_3)$. This
substitution is made in all contributions to ${\rm d}\sigma^2$, including
potential cross-terms such as ${\rm d}x_1{\rm d}x_2$ as well as in
position-dependent metric components. In this way, a general curved
4-dimensional line element can be treated without any restrictions.

Once a suitable auxiliary spacetime has been identified, the ingredients used
in our construction, given by parallel and central projection as well as light
cones and planes of constant $\zeta$, can be generalized using geodesics as
well as the causal structure of a curved geometry. For instance, we replace
the constant-$\zeta$ plane in which we map hyperbolae to circles, by a plane
defined by geodesics starting at the same point of the $\zeta$-axis with
initial tangent vectors $v^{\mu}$ such that
$v^{\mu}\partial_{\mu}\zeta=0$. These geodesics might intersect at some
distance from the $\zeta$-axis and fail to form a plane there, but locally
around this axis they describe an open region on a deformed plane. Our
projections can therefore be constructed locally, in sufficiently small
neighborhoods around points in the original spacetime.

The same geodesics as used to define the local plane then provide the curves
along which we perform central projections from hyperbolae to
circles. Hyperbolae are now defined as intersections of the geodesic plane
with light cones in the auxiliary spacetime. Generalized circles are obtained
after we flip $g_{tt}$ with $g_{\zeta\zeta}$ in timelike regions, using
Gaussian normal coordinates. In spacelike regions, we flip the entire spatial
contributions in the auxiliary spacetime metric.

This outline includes all the steps of our construction. There is therefore a
good potential that our constructions can be generalized to generic curved
spacetime backgrounds as well as self-interacting gravity, but details remain
to be worked out.

\section{Conclusions}

Many constructions in quantum field theories for fundamental physics on
spacetime require different kinds of regularization, among which the Wick
rotation plays an important role. However, it introduces complex time and
therefore makes it hard to develop an intuitive understanding. It has also been
known for some time to fail in important situations such as quantum field
theories at constant chemical potential, relevant in particular in studies of
the quark-gluon plasma in lattice QCD, and for quantum field theories on
non-stationary curved backgrounds or for quantum gravity. In fact, most
approaches to quantum gravity make essential use of a Wick rotation without
being able to provide an independent justification.

These questions motivate us to look for alternative constructions that may
fulfill the same purpose, but be applicable more generally. We have therefore
developed a geometrical mapping between Minkowski spacetime and Euclidean
space which, by its very nature, is not an isometry of the metric structures
but relates them by implementing one of the implicit regularization features
of the Wick rotation. In particular, the mapping relates non-compact
hyperbolae to compact circle segments in such a way that the standard finite
values are obtained in QFT-type integrals, but without using complex numbers
in the metric transformation. Complex numbers may however appear in some
expressions derived from the metric, such as the transformed
$\sqrt{-\det(\eta_{\mu\nu})}$ after the Minkowski metric $\eta_{\mu\nu}$ is
replaced by a Euclidean version. Keeping the metric real, the geometrical
nature of our transformation implies a more intuitive understanding of this
regularization step. Another ingredient of our transformation is an equal
treatment of space and time, both of which may change sign in a line element
depending on whether we consider a timelike or spacelike region relative to
the origin. Our construction therefore combines a standard Wick rotation with
an alternative suggested in \cite{NewWick} for cosmological spacetimes, while
removing any reference to complex numbers in the transformation.

Moreover, we showed that the same construction can be used to transform
Minkowskian action principles to Euclidean ones, replacing complex transition
amplitudes in a path integral with real statistical weights in a state
sum. Compared with a Wick rotation, there are two key advantages in that (i)
field theories at constant chemical potential can be treated in a real manner
on the Euclidean side, for both bosonic and, with an additional ingredients
from the older \cite{FermionWick}, fermionic ones, and (ii) non-stationary
curved backgrounds and gravitational actions can, at least in principle, be
analyzed using similar mappings generalized to curved spacetime. While much
remains to be worked out in these significant applications, we believe that
our examples have demonstrated early promise. 

\section*{Acknowledgements}

This work was supported in part by NSF grants PHY-2206591 (M.B.) and the
Wilson Faculty Fellowship awarded to V.G. by the Pennsylvania State
University.

\begin{appendix}

\section{Connection to the RBS construction} \label{sec:RBS}

In \cite{RBS}, a Euclideanization procedure for the MS was proposed. Dubbed
the Renormalized Blended Spacetime, or RBS, it consisted of two steps:
blending and ``renormalization'' (not to be confused with renormalization in
quantum field theory). Consider two inertial observers, one labelled
``unprimed'' who is stationary, and the other labelled ``primed'' who is
moving at a constant speed $v$ in the $+\hat{\textbf{x}}$ direction relative
to the unprimed observer. They both observe an event whose coordinates they
note down as $(x_0, ct_0)$ (unprimed observer) and $(x_0', ct_0')$ (primed
observer). Let the two inertial observers adopt each other's time coordinates
of an event as their own, while retaining their original space coordinates of
that event; the observers become ``blended''. Invariance of the Minkowski line
element then implies 
\begin{equation} 
  x_0^2+\left (ct_0' \right )^2 = \left ( x_0' \right )^2+(ct_0)^2\,.
\end{equation}
Using hyperbolic coordinates $(\xi_0,\sigma_0)$ of the event $(x_0,ct_0)$ as
defined in Eq.~(\ref{eqn:p=xi(cosh_sigma, sinh_sigma)}), the results of
\cite{RBS} can be expressed as
\begin{equation} \label{eqn:blending}
  x_0^2+\left (ct_0' \right )^2 = \left ( x_0' \right )^2+(ct_0)^2=\xi_0^2\chi^2
\end{equation}
with
\begin{equation}
  \chi^2 = \cosh{\alpha} \cosh{(2\sigma_0-\alpha)}
\end{equation}
and
\begin{equation}
    \cosh{\alpha} = \frac{1}{\sqrt{1-v^2/c^2}}\,.
\end{equation} 

As suggested by the signs in (\ref{eqn:blending}), this process reformulates
the Lorentz boosts into Euclidean rotations while retaining the original
spacetime hyperbola (defined by the constant $\xi_0$) describing worldlines of
constant spacetime length from the origin. In the next step, by
``renormalizing'' the blended coordinates with $\chi$ that is a function of the
relative velocity, $v$, between the two frames,
\begin{align} \label{eqn:renormalization}
    \Bar{x}_0 &=x/\chi, \hspace{5mm} \Bar{t}_0 = t_0/ \chi \\
    \Bar{x}_0' &=x'/\chi, \hspace{5mm} \Bar{t}_0' = t_0'/ \chi
\end{align}
the hyperbola is transformed
into a Euclidean circle of constant radius $\xi_0$.
With these two steps, one obtains the RBS coordinates complete with new light
lines \cite{RBS}, but now with a Euclidean construction: 
\begin{equation} \label{eqn:RBS-circle}
    \Bar{x}_0^2+(c \Bar{t}_0')^2 = \Bar{x}_0'^2+(c\Bar{t}_0)^2 = \xi_0^2\,.
  \end{equation}
  
We now note that for the special case of $v=0$, the angle $\alpha=0$ and the
two inertial frames coincide, namely, $x_0 \rightarrow x_0'$ and $t_0
\rightarrow t_0'$. Further, $\chi=\cosh{2\sigma_0}=1/\Omega$ (see
Eq.~(\ref{eqn:Scaling_Factor_Omega_as_a_f(xo, to)})), and the Euclidean
coordinates become 
\begin{equation} \label{eqn:RBS-v=0}
    \Bar{x}_0 =\Omega x_0, \hspace{5mm} \Bar{t}_0 = \Omega t_0\\
\end{equation}
Thus the point transformation for the special case of $v=0$ in the RBS
transformation coincides with Eq.~(\ref{eqn:Complete-Projection-p0-to-pBar}): 
\begin{equation}\label{eqn:point transformation-RBS}
    p_0 =(x_0, ct_0) \rightarrow  \Bar{p}= (\Omega x_0, \Omega ct_0)
\end{equation}
For the points $p_0$ on the light lines, the discussion of the above
transformation is a bit subtle (see Sec ~\ref{sec:Lightcone}) and is discussed
in detail in \cite{RBS}. Other possible cuts of the $xt\zeta$ auxillary space
in Fig.~\ref{fig:conicproj} correspond to other limiting cases in the
RBS. This discussion is left for future explorations.
\end{appendix}


\begin{thebibliography}{3}

\bibitem{QFT}
Srednicki, M.,
\textit{Quantum Field Theory}.
2006,
http://www.physics.ucsb.edu/\~{}mark/qft.html.

\bibitem{Instantons}
Coleman, S.,
\textit{The Uses of Instantons}.
1979,
https://doi.org/10.1007/978-1-4684-0991-8\_16

\bibitem{SignProblem}
  Gattringer, C.\ and Langfeld, K., Approaches to the sign problem in lattice
  field theory, Int.\ J.\ Mod.\ Phys.\ A 31 (2016) 1643007, arXiv:1603.09517
  
\bibitem{FiniteDensityQCD}
  Nagata, K., Finite-density lattice QCD and sign problem:
current status and open problems, Prog.\ Part.\ Nucl.\ Phys.\ 127 (2022)
103991, arXiv:2108.12423

\bibitem{FermionWick}
  van Nieuwenhuizen, P.\ and Waldron, A., On Euclidean spinors and Wick
  rotations, Phys.\ Lett.\ B 389 (1996) 29--36, hep-th/9608174

\bibitem{RBS}
Gopalan, V.,
Relativistic spacetime crystals.
(2021). 
\textit{Acta Crystallographica Section A: Foundations and Advances}, 77, 242–256. https://doi.org/10.1107/S2053273321003259.

\bibitem{Bohr}
  Halvorson, H., Complementarity of representations in quantum mechanics, quant-ph/0110102

\bibitem{EuclideanSpinor}
  Wetterich, C., Spinors in euclidean field theory,
complex structures and discrete
symmetries, Nuc.\ Phys.\ B 852 (2011) 174--234, arXiv:1002.3556

\bibitem{EuclideanQED}
  Schwinger, J., Euclidean Quantum Electrodynamics, Phys.\ Rev.\ 115 (1959) 721

\bibitem{NewWick}
  Gutowski, J., Mohaupt, T.\ and Pope, G., Cosmological Solutions, a New
  Wick-Rotation, and the First Law of Thermodynamics, 	arXiv:2008.06929

  
\end{thebibliography}
\end{document}